\documentclass[sigconf, nonacm]{acmart}

\newcommand\vldbdoi{10.14778/3476249.3476253}
\newcommand\vldbpages{1937 - 1949}
\newcommand\vldbvolume{14}
\newcommand\vldbissue{11}
\newcommand\vldbyear{2021}
\newcommand\vldbauthors{\authors}
\newcommand\vldbtitle{\shorttitle} 
\newcommand\vldbavailabilityurl{https://github.com/NAIST-SE/PATSQL}

\newcommand\vldbpagestyle{empty} 

\usepackage{color}
\usepackage[noend]{algorithmic}
\usepackage{soul} 
\usepackage{booktabs}
\usepackage[a-2b]{pdfx}

\newcommand\added[1]{#1} 
\newcommand\minor[1]{#1} 

\begin{document}
\title{PATSQL: Efficient Synthesis of SQL Queries from Example Tables with Quick Inference of Projected Columns}


\author{Keita Takenouchi}
\affiliation{%
  \institution{NTT DATA}
  \city{Tokyo}
  \country{Japan}}
\email{Keita.Takenouchi@nttdata.com}

\author{Takashi Ishio}
\affiliation{%
  \institution{Nara Institute of Science and Technology}
  \city{Nara}
  \country{Japan}}
\email{ishio@is.naist.jp}

\author{Joji Okada}
\affiliation{%
  \institution{NTT DATA}
  \city{Tokyo}
  \country{Japan}}
\email{Joji.Okada@nttdata.com}

\author{Yuji Sakata}
\affiliation{%
  \institution{NTT DATA}
  \city{Tokyo}
  \country{Japan}}
\email{Yuji.Sakata@nttdata.com}

\def\tool{P{\footnotesize AT}SQL}
\def\toolf{P{\footnotesize AT}SQL$_\text{5}$}
\def\scythe{S{\footnotesize CYTHE}}
\def\morpheus{M{\footnotesize ORPHEUS}}
\def\vsk{\textit{s}}
\def\vpg{\textit{p}}
\def\vsset{\textit{S}}
\def\vpset{\textit{P}}

\begin{abstract}
SQL is one of the most popular tools for data analysis, and it is now used by an increasing number of users without having expertise in databases. Several studies have proposed programming-by-example approaches to help such non-experts to write correct SQL queries. 
While existing methods support a variety of SQL features such as aggregation and nested query, they suffer a significant increase in computational cost as the scale of example tables increases.
In this paper, we propose an efficient algorithm utilizing properties known in relational algebra to synthesize SQL queries from input and output tables. 
Our key insight is that a projection operator in a program sketch can be lifted above other operators by applying transformation rules in relational algebra, while preserving the semantics of the  program.  
This enables a quick inference of appropriate columns in the projection operator, which is an essential component in synthesis but causes combinatorial explosions in prior work. 
We also introduce a novel form of constraints and its top-down propagation mechanism for efficient sketch completion. 
We implemented this algorithm in our tool \tool\ and evaluated it on \added{226} queries from prior benchmarks and Kaggle's tutorials. As a result, \tool\ solved \added{68}\% of the benchmarks and found \added{89}\% of the solutions within a second.
Our tool is available at \url{https://naist-se.github.io/patsql/}.
\end{abstract}

\maketitle

\pagestyle{\vldbpagestyle}
\begingroup\small\noindent\raggedright\textbf{PVLDB Reference Format:}\\
\vldbauthors. \vldbtitle. PVLDB, \vldbvolume(\vldbissue): \vldbpages, \vldbyear.\\
\href{https://doi.org/\vldbdoi}{doi:\vldbdoi}
\endgroup
\begingroup
\renewcommand\thefootnote{}\footnote{\noindent
This work is licensed under the Creative Commons BY-NC-ND 4.0 International License. Visit \url{https://creativecommons.org/licenses/by-nc-nd/4.0/} to view a copy of this license. For any use beyond those covered by this license, obtain permission by emailing \href{mailto:info@vldb.org}{info@vldb.org}. Copyright is held by the owner/author(s). Publication rights licensed to the VLDB Endowment. \\
\raggedright Proceedings of the VLDB Endowment, Vol. \vldbvolume, No. \vldbissue\ %
ISSN 2150-8097. \\
\href{https://doi.org/\vldbdoi}{doi:\vldbdoi} \\
}\addtocounter{footnote}{-1}\endgroup

\ifdefempty{\vldbavailabilityurl}{}{
\vspace{.3cm}
\begingroup\small\noindent\raggedright\textbf{PVLDB Artifact Availability:}\\
The source code, data, and/or other artifacts have been made available at \url{\vldbavailabilityurl}.
\endgroup
}


\begin{figure*}[t]
    \centering
    \includegraphics[width=0.95\linewidth]{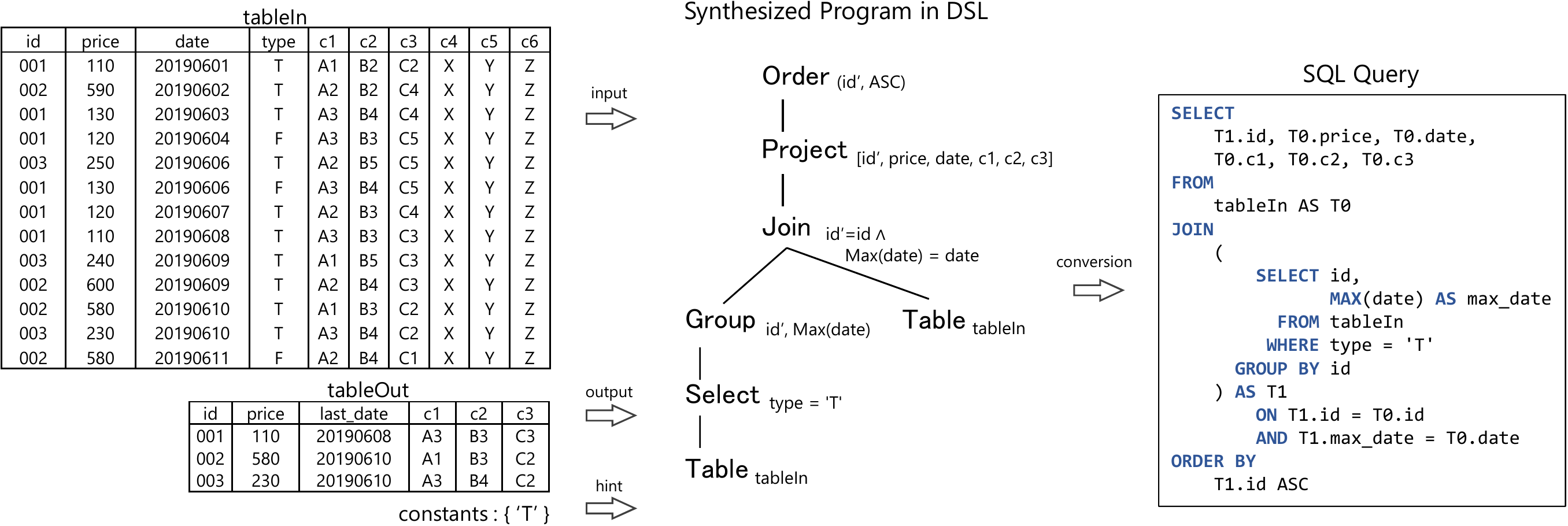}
    \caption{The overview of our approach. It synthesizes a program in our DSL, and converts it into an SQL query.}
    \label{fig:overview}
\end{figure*}%

\section{Introduction}
SQL is a query language that manages data in relational databases, and it is commonly used in a wide range of software systems ranging from web applications to banking systems. With the growing popularity of data analysis in recent years, an increasing number of users without having expertise in databases are using SQL \cite{tse_data}. However, it is not easy for such non-experts to write SQL queries since they need to express a variety of analytical needs. 

Programming-by-example (PBE) is a program synthesis technique that automatically synthesizes programs from input and output (I/O) examples. PBE is known to be a practical approach to help non-experts to implement programs \cite{microsoft-survey, pbe}. In the context of SQL, queries are automatically synthesized from I/O tables that the user provides as an example.
In recent years, several studies have proposed techniques to automatically synthesize SQL queries \cite{scythe,sql-synthesizer,sqlsol, squares} or table manipulation programs \cite{morpheus, autopandas} from I/O tables. 
\scythe\ \cite{scythe} synthesizes SQL queries that support highly expressive features such as projection, join, grouping, aggregation and union. \scythe\ even supports the synthesis of nested queries, which have other queries inside. These features enable users to obtain practical queries that gain insights from accumulated data.

However, these methods commonly have serious performance issues depending on the scale of I/O tables. The reason is that these algorithms need to enumerate a large number of candidates for each part of a program. 
For example, \scythe\ suffers an exponential increase in computational cost as the number of columns increases since it needs to enumerate all the permutations of them. However, the schemas of tables in real-world databases are not small, and thus the user fails to obtain queries in most of the practical scenarios.

In this paper, we propose a sketch-based algorithm that synthesizes SQL queries from I/O tables. 
\added{
The illustrative example is shown in Figure~\ref{fig:overview}. 
}
Our algorithm is efficient in terms of the execution time and the scale of supported tables. While it requires fewer hints (i.e. constants used in a query) than in prior work~\cite{scythe}, it maintains the high expressiveness of synthesized queries including aggregations, nested queries and window functions. To the best of our knowledge, this is the first SQL synthesizer that supports window functions such as cumulative sum and rank of each row, which have gained more popularity in database communities over the past years. Note that our algorithm does not depend on the column names because they do not necessarily indicate the correspondence between input and output columns.

To achieve the efficiency, we integrate properties known in relational algebra into sketch-based program synthesis. Our key insight is that the projection operator \added{(i.e. \texttt{Select} keyword in SQL)} in a program sketch can be lifted above other operators by applying transformation rules in relational algebra without changing the semantics of the program. 
By leveraging this insight, we can avoid a combinatorial explosion that existing methods suffer during the completion of the projected columns.

We implemented this algorithm in our tool \tool. 
To evaluate it, we used 193 SQL queries from Stack~Overflow and a textbook on databases, as in prior work. We also collected 33 queries from Kaggle's tutorials, which deal with real table schemas used for data analysis. The result shows \tool\ significantly outperforms a state-of-the-art method \scythe\ in terms of the execution time and the scalability of I/O tables it can handle. In particular, \tool\ solved \added{68\%} of the benchmarks while \scythe\ solved \added{57\%} of them. Moreover, \tool\ found \added{136} solutions (\added{89\%} of the solved benchmarks) within a second while \scythe\ found only 28 solutions. 

The main contributions of this paper are as follows.
\begin{itemize}
    \item We propose a novel technique that synthesizes SQL queries from I/O tables. It has strengths in both the execution time and the scale of I/O tables it can handle. It also supports the high expressiveness of synthesized queries including grouping, aggregation, nested query and window functions only with hints about used constants. 
    \item \added{We describe the synthesis algorithm that focuses on the efficient completion of the projected columns by leveraging properties known in relational algebra.} 
    \item We propose a user interface that is inspired by the concept of live programming. The user can get real-time feedback on the synthesized queries each time s/he updates tables.
    \item We implement the algorithm in our tool \tool\ and evaluate it on various benchmarks including queries that we collected from Kaggle's tutorials. The results show that \tool\ significantly outperforms a state-of-the-art algorithm \scythe\ in terms of the execution time and the scalability of I/O tables. 
\end{itemize}

\section{Overview}
In this section, we provide an overview of our approach with an illustrative example and define the problem we solve in this paper. 

\subsection{Illustrative Example}
We show an example to illustrate the usefulness of our PBE method. Suppose that a non-expert user wants to extract data from a relational database. A table named \texttt{tableIn} keeps track of the price history of items, and the schema consists of \emph{item ID}, \emph{price}, \emph{updated date}, \emph{type}, and other six columns \emph{c1}$,\dots,$\emph{c6}. Now the user wants to know \emph{the latest price and the updated date along with the values in c1, c2 and c3 for each item whose type is ``T''}, but has difficulty writing  such a query since  the user is not familiar with the syntax and semantics of SQL.

Instead of writing the query from scratch, the user can use our tool \tool. First, the user gives an example of the I/O tables that should be satisfied by the query.  The user also needs to provide hints for synthesis, i.e., the constants used in predicates in the query. The I/O tables and hint are shown on the left of Figure \ref{fig:overview}. Here \texttt{tableIn} and \texttt{tableOut} are the input and output tables, respectively. The constant value ``\texttt{T}'' is given as a hint.
Then, \tool\ synthesizes a program in our DSL 
(see Section \ref{sec:dsl} for the details)
that satisfies the I/O tables and makes use of the hints provided. 
In this case, \tool\ synthesizes the program shown in the center of Figure \ref{fig:overview}. This program first filters out the records that do not belong to the type ``\texttt{T}'' from the input table, and it calculates the maximum date for each item. Then, it joins the aggregation result and the input table with two key pairs, and it extracts the desired columns and sorts records.
Finally, \tool\ converts the synthesized program in DSL into a SQL query and returns it as a result. The synthesized SQL query is shown on the right of Figure \ref{fig:overview}. The user can extract the desired records by executing it on the original table. Since the query has a nested structure having  aggregation with grouping, it is not easy for the user to write it without the aid of \tool. Note that the conversion from our DSL to SQL is straightforward because our DSL is based on relational algebra, which is a theoretical foundation for SQL. Hence, we omit its details in this paper and focus on the synthesis algorithm in our DSL.

In general, the schemas of example tables given to a PBE tool should be the same as the schemas of the original tables stored in a database. This is because queries synthesized from simplified tables can cause SQL errors when it is executed in the database. 
For example, suppose the user deletes the columns c1, c2 and c3 from the output \minor{example in Figure~\ref{fig:overview}} because the user mistakenly thinks these columns do not affect the structure of a synthesized query. Then, a PBE tool may well return the following query, which is much simpler than the desired query in Figure \ref{fig:overview}.
\begin{verbatim}
    SELECT id, min(price), max(date)
    FROM tableIn WHERE type = 'T' GROUP BY id
\end{verbatim}
\minor{After} obtaining this query, the user adds the columns c1, c2 and c3, the values of which the user wants to see, and then the user creates the following query. 
\begin{verbatim}
    SELECT id, min(price), max(date), c1, c2, c3
    FROM tableIn WHERE type = 'T' GROUP BY id
\end{verbatim}
However, an SQL error occurs when the query is executed \minor{in the database} because the columns c1, c2 and c3 are not included in the grouping key. To avoid such undesired situations, we assume that \minor{example} tables given to our \minor{tool} preserve the original schemas.

\def\term{\texttt}
\def\var{\textit}

\begin{figure}[t]
{\small
   \begin{align*}
    \langle \textit{table} \rangle ::=\ &\ \term{Table}(\var{tname}) \\ 
               \mid\ &\ \term{Order}(\langle \textit{table} \rangle, [\langle \textit{key} \rangle_1, \dots, \langle \textit{key} \rangle_n]) \\
               \mid\ &\ \term{Distinct}(\langle \textit{table} \rangle) \\
               \mid\ &\ \term{Project}(\langle \textit{table} \rangle, [\langle \textit{col} \rangle_1,\dots, \langle \textit{col} \rangle_n]) \\
               \mid\ &\ \term{Select}(\langle \textit{table} \rangle, \langle \textit{pred} \rangle) \\
               \mid\ &\ \term{Group}(\langle \textit{table} \rangle, [\textit{cname}_1, \dots, \textit{cname}_m], [\langle \textit{gc} \rangle_1, \dots, \langle \textit{gc} \rangle_n]) \\
               \mid\ &\ \term{Window}(\langle \textit{table} \rangle, [\langle \textit{win} \rangle_1, \dots, \langle \textit{win} \rangle_n])\\
               \mid\ &\ \term{Join}(\langle \textit{table} \rangle, \langle \textit{table} \rangle, \langle \textit{pairs} \rangle) \\
               \mid\ &\ \term{LeftJoin}(\langle \textit{table} \rangle, \langle \textit{table} \rangle, \langle \textit{pair} \rangle) \\
     \langle \textit{key} \rangle  ::=\ &\ (\langle \textit{col} \rangle, \term{Asc}) \mid (\langle \textit{col} \rangle, \term{Desc}) \\
      \langle \textit{col} \rangle ::=\ &\ \var{cname} \mid \langle \textit{gc} \rangle \\
      \langle \textit{gc} \rangle ::=\ &\  \langle \textit{agg} \rangle ( \langle \textit{col} \rangle ) \\
     \langle \textit{win} \rangle ::=\ &\ ( \langle \textit{func} \rangle, col, [\textit{cname}_1, \dots, \textit{cname}_m], \langle \textit{key} \rangle) \\
     \langle \textit{pairs} \rangle ::=\ &\ \langle \textit{pair} \rangle \wedge \cdots \wedge  \langle \textit{pair} \rangle \\
     \langle \textit{pair} \rangle ::=\ &\ \langle \textit{col} \rangle\ \term{=}\ \langle \textit{col} \rangle \\
     \langle \textit{pred} \rangle ::=\ &\ \langle \textit{clause} \rangle \vee \cdots \vee \langle \textit{clause} \rangle  \\
     \langle \textit{clause} \rangle ::=\ &\ \langle \textit{prim} \rangle \wedge \dots \wedge \langle \textit{prim} \rangle \\
     \langle \textit{prim} \rangle ::=\ &\  \langle \textit{col} \rangle\ \langle \textit{binop} \rangle\ \var{const}\ \mid
                    \term{IsNull}(\var{cname}) \mid
                    \term{IsNotNull}(\var{cname}) \\
      \langle \textit{agg} \rangle ::=\ &\ \term{Max} 
               \mid \term{Min} 
               \mid \term{Count} 
               \mid \term{Sum}
               \mid \term{Avg}
               \mid \term{CountDistinct}\\ 
               \mid\ &\ \term{ConcatComma} 
               \mid \term{ConcatSpace}
               \mid \term{ConcatSlash} \\
     \langle \textit{func} \rangle ::=\ &\ \term{Max} 
               \mid \term{Min} 
               \mid \term{Count} 
               \mid \term{Sum} 
               \mid \term{Rank} \\
     \langle \textit{binop} \rangle ::=\ &\
               \term{=} \mid 
               \term{<} \mid 
               \term{<=} \mid 
               \term{>} \mid 
               \term{>=} \mid 
               \term{<>} 
     \end{align*}
}
    \caption{
    The grammar of our DSL. \var{tname} denotes the name of a input table. \var{cname} denotes a column name. \var{const} denotes a constant value. $n$ is the size of a vector, and $m$ is the size of grouping keys, which is limited to two or less.
    }
    \label{fig:grammar}
\end{figure}%

\subsection{Problem Definition}
The input of our algorithm is a tuple $(\mathcal{E}, \mathcal{C})$, where $\mathcal{E} = (\vec{T}_\text{in}, T_\text{out})$ is an example of input tables $\vec{T}_\text{in}$ and an output table $T_\text{out}$, and  $\mathcal{C} = \{v_1, \ldots, v_k\}$ is a set of typed constants. The schema of each column has a name and a type. The types consist of  \texttt{Str}, \texttt{Int}, \texttt{Dbl} and \texttt{Date}. In this paper, we propose an algorithm that takes $(\mathcal{E}, \mathcal{C})$ as input, and returns a program $\vpg$ in our DSL that satisfies $\vpg(\vec{T}_\text{in}) = T_\text{out}$, where the constants used in $\vpg$ are included in $\mathcal{C}$. 
\added{
Here the equality operator ($=$) compares two tables by treating records as a list if $T_\text{out}$ has at least one sorted column, and treating them as a multiset otherwise. That is, we synthesize programs with sort operators whenever possible. Of course, this policy possibly synthesizes extra sort operators, but it is important to support the sort operator since \texttt{ORDER}~\texttt{BY} clause is often used in real-world SQL queries~\cite{sql-synthesizer}.
}

The limitation of asking the user to provide the constants is the same as that of \scythe \ \cite{scythe}. The constants are used not only for improving the synthesis performance but for directly reflecting the user's intention. We believe this limitation is not an obstacle in practice because it is known that users asking SQL questions on Stack~Overflow are usually ready to provide such constants even when they do not know how to write correct queries \cite{scythe}. In addition to constants, \scythe\ requires hints about aggregation functions while \tool\ does not. One of the \tool's advantages is that it works with a more limited kind of user hints than in prior work~\cite{scythe, squares}. 

\section{Domain-Specific Language}
\label{sec:dsl}

\added{
In this section, we describe our domain-specific language (DSL) that determines the search space of synthesis problems. The DSL is a kind of extended relational algebra with additional operators such as window functions.
} 
Before describing each operator in our DSL, we emphasize that DSLs for program synthesis need to be carefully designed with the trade-off between expressiveness and efficiency of synthesis \cite{flashfill, mitra}. Following this policy, our DSL is designed to support as many SQL features as possible while allowing the synthesis algorithm to leverage properties of relational algebra. 

Figure~\ref{fig:grammar} shows the grammar, and the start symbol is $\langle \textit{table} \rangle$. 
We refer to the elements other than \term{Table} in the right-hand side of the rule $\langle \textit{table} \rangle$ as \textit{operators}. Each operator is a function that takes one or more tables as input and returns a table. 

The semantics of the operators is as follows. Here we use a vector such as $\vec{c}$ to represent multiple elements. 
$\term{Project}(T, \vec{\textit{c}})$ extracts columns $\vec{\textit{c}}$ from table $T$.
$\term{Select}(T, \textit{pred})$ selects rows in $T$ that satisfy a predicate \textit{pred}. This predicate is in a conjunctive normal form, namely an AND of ORs.
$\term{Group}(T, \vec{\textit{c}}, \vec{\textit{gc}})$ groups rows in $T$ by keys $\vec{c}$ and returns a new table, whose columns consist of the keys $\vec{c}$ and aggregation results $\vec{\textit{gc}}$. 
$\term{Window}(T, \vec{\textit{w}})$ appends the resulting columns of window functions $\vec{\textit{w}}$ to $T$. Each column $w$ consists of a window function, a target column, partitioning keys and a sort key. 
$\term{Join}(T_1, T_2, \textit{pairs})$ executes inner join on  $T_1$ and  $T_2$ with key \textit{pairs}.
Likewise $\term{LeftJoin}(T_1, T_2, pair)$ executes left join with a key \textit{pair}.
$\term{Distinct}(T)$ removes duplicated rows in $T$.
$\term{Order}(T, \vec{\textit{key}})$ sorts rows in $T$ according to key columns and directions  $\vec{\textit{key}}$.

Our DSL supports a variety of SQL features used in practice. It supports \texttt{SELECT}, \texttt{WHERE}, \texttt{GROUP}~\texttt{BY}, \texttt{JOIN}, \texttt{LEFT}~\texttt{JOIN}, \texttt{DISTINCT}, \texttt{ORDER}~\texttt{BY} and  \texttt{HAVING}. It also supports operators that can be expressed by other operators such as \texttt{EXISTS} (expressed by \texttt{JOIN}), \texttt{NOT}~\texttt{EXISTS} (expressed by \texttt{LEFT}~\texttt{JOIN} and \texttt{IS}~\texttt{NULL} \cite{trans}), and  \texttt{RIGHT} \texttt{JOIN} (expressed by \texttt{LEFT}~\texttt{JOIN}). \texttt{BETWEEN} and \texttt{IN} in predicates can also be rewritten using \texttt{OR}. In summary, our DSL supports 17 out of 20 keywords that are most popular among SQL users according to the questionnaire survey in 2013 \cite{sql-synthesizer}. Besides, it supports window functions and nested queries since DSL operators can have other operators as their children. 
On the other hand, we do not support \texttt{UNION} operator. We discuss this limitation in Section \ref{sec:adapt}. 

\begin{figure}[t]
{\small
   \begin{align*}
   \langle \textit{s} \rangle ::=\ &\  \term{Table}(\square) 
               \mid\    \term{Order}(\langle  \textit{s} \rangle, \square) 
               \mid\    \term{Distinct}(\langle  \textit{s} \rangle) \\
               \mid\ &\ \term{Project}(\langle  \textit{s} \rangle, \square)  
               \mid\    \term{Select}(\langle  \textit{s} \rangle, \square) 
               \mid\    \term{Group}(\langle  \textit{s} \rangle, \square, \square) \\
               \mid\ &\ \term{Window}(\langle  \textit{s} \rangle, \square)
               \mid\    \term{Join}(\langle  \textit{s} \rangle, \langle  \textit{s} \rangle, \square) 
               \mid\    \term{LeftJoin}(\langle  \textit{s} \rangle, \langle  \textit{s} \rangle, \square) 
     \end{align*}
}
    \caption{
    The grammar of the sketches in our DSL. The start symbol is $\langle s \rangle$.The symbol `$\square$' denotes an uninstantiated part.
    }
    \label{fig:abs_grammar}
\end{figure}%

\section{Preliminaries}
In this section, we introduce key concepts that are used in our synthesis algorithm.

\subsection{Sketch and its Completion} 
\label{sec:pre_sketch}

We refer to a program with uninstantiated parts as \textit{sketch}, and an uninstantiated part is denoted as the symbol `$\square$'. Specifically, the sketches in our algorithm are the languages that are produced by the grammar of Figure \ref{fig:abs_grammar}. In the syntax tree of a sketch, a leaf node corresponds to a \texttt{Table} while a non-leaf node corresponds to an operator. We call \textit{sketch $\vsk'$ is a child of sketch $\vsk$} when the node for $\vsk'$ is a child of \minor{the node} for $\vsk$ in the corresponding syntax tree. Also, \textit{sketch completion} is an action that fills `$\square$'s with concrete structures. Our synthesis algorithm generates sketches and tries to complete each of them. Creating a sketch determines the operators used in the resulting program and enables us to validate its structure. 
\added{ 
Here we define the size of a sketch as follows. The size of the operators other than \texttt{Window} is one. The size of \texttt{Window} is two because the functionality is more complicated than the other operators, namely partitioning cells, applying window functions, and appending the results to the original table. We then define $\textsc{Size}(s)$ function as the sum of each operator's size in a sketch $s$.  
}

\added{ 
For instance, the illustrative example in Figure~\ref{fig:overview} employs
}
the sketch $s =$
\texttt{Order}(\texttt{Project}(\texttt{Join}(\texttt{Group}(\texttt{Select}(\texttt{Table}($\square$), $\square$), $\square$), \texttt{Table}($\square$),$\square$), $\square$), $\square$), and  \added{$\textsc{Size}(s)$ returns five}. The results of sketch completion of $s$ include the program in the center of Figure~\ref{fig:overview}. 

\begin{figure}[t]
   \begin{align*}
\varphi(T, T') ::=\ & \Leftrightarrow_{R}(T, T')\ |\ \Mapsto_{R}(T,T')\ |\ \top \\
R ::=\ & =_\text{bag}\ |\ \subseteq_\text{bag}\ |\ \ =_\text{set}\ |\ \subseteq_\text{set}\\
\Leftrightarrow_{R}(T, T') \leftrightarrow\ & |T| = |T'| \land \forall i \in [1, |T|].\ R(T[i], T'[i]) \\
\Mapsto_{R}(T,T') \leftrightarrow\ & \forall i \in [1, |T|].\ \exists i' \in [1, |T'|].\ R(T[i], T'[i']) 
   \end{align*}
    \caption{
      The definition of our table relation $\varphi$. \textit{T} and \textit{T'} refer to tables. $|T|$ is the number of columns in $T$, and $T[i]$ is the i-th column in $T$.
    }
    \label{fig:constraint}
\end{figure}

\begin{figure}[t]
    \centering
    \includegraphics[width=0.5\linewidth]{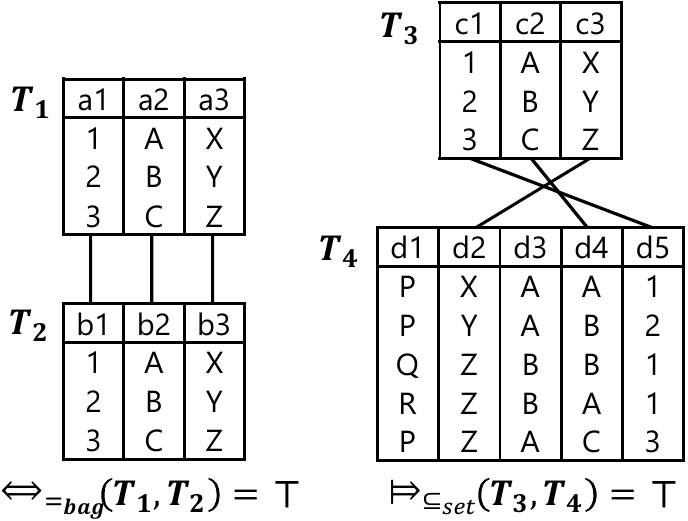}
    \caption{Examples of our table relation $\varphi$. The edges between columns represent the column correspondence that is needed to hold a relation $\varphi$.}
    \label{fig:example_of_constraint}
\end{figure}%

\subsection{Table Inclusion Relation $\varphi$}
In our synthesis algorithm, the knowledge of the output table is propagated in the form of our original constraint. The constraint denoted as $\varphi(T, T')$ is an inclusion relation between two tables $T$ and $T'$, and it returns true ($\top$) or false ($\bot$). Figure \ref{fig:constraint} shows the definition. The relation $\varphi$ is in one of the three states: $\Leftrightarrow$, $\Mapsto$ and $\top$. The relation $\Leftrightarrow_{R}(T, T')$ means that the columns in $T$ and $T'$ correspond to each other in their order. $\Mapsto_{R}(T,T')$ means that there exist at least one column in $T'$ that correspond to each column in $T$. 
\added{ 
Specifically, we calculated the column mapping by enumerating the column pairs of the two tables. 
}
The parameter $R$ in $\Leftrightarrow_{R}$ and $\Mapsto_{R}$ is a binary relation between columns, and holds a type of comparison: equality ($=$) or inclusion ($\subseteq$), and a treatment of cells: multiset (bag) or set (set).

We show examples of these relations in Figure \ref{fig:example_of_constraint}. By using the tables $T_1$ and $T_2$, the predicate $\Leftrightarrow_{=_\text{bag}}(T_1, T_2) = \top$ holds since the column relations $=_\text{bag}(\text{a1}, \text{b1})$, $=_\text{bag}(\text{a2}, \text{b2})$ and $=_\text{bag}(\text{a3}, \text{b3})$ are true, and the columns correspond to each other in their order. On the other hand, the relation $\Mapsto_{\subseteq_\text{set}}(T_3, T_4) = \top$ holds since the column relations $\subseteq_\text{set}(\text{c1}, \text{d5})$, $\subseteq_\text{set}(\text{c2}, \text{d4})$ and $\subseteq_\text{set}(\text{c3}, \text{d2})$ are true, and there exists a column in $T_4$ that corresponds to each column in $T_3$.

The constraint $\varphi$ can be represented as a tuple $(M, R)$ except for the case of $\top$, where $M \in \{ \Leftrightarrow$, $\Mapsto \}$ and $R \in \{ =_\text{bag}, \subseteq_\text{bag}, =_\text{set}, \subseteq_\text{set}\}$. In the rest of this paper, we denote $\varphi(T, T')$ as $\varphi = (M, R)$ for short when the tables $T$ and $T'$ are obvious from the context. For example, we use notations such as $\varphi = (\Leftrightarrow, =_\text{bag})$ or $\varphi = (\Mapsto, \subseteq_\text{set})$.

\setuldepth{ABC}

\begin{figure}
    \begin{flushleft}
        \ul{\textsc{Synthesize}}$(\mathcal{E}, \mathcal{C})$\\
        {\small
        \textbf{Input}\ $\mathcal{E} = (\vec{T}_\text{in}, T_\text{out})$:\ input and output tables \\
        \ \ \ \ \ \ \ \ \ \ \ $\mathcal{C}$:\ constants used in predicates \\
        \textbf{Output} a synthesized program}
    \end{flushleft}
    \begin{algorithmic}[1]
    \IF{$\textsc{IsSorted}(T_\text{out})$}
        \STATE $\vsset \leftarrow \{ \texttt{Table}(\square) \}$
    \ELSE
        \STATE $\vsset \leftarrow \{ \texttt{Order}(\texttt{Table}(\square), \square) \}$
    \ENDIF
    \WHILE{true}
        \STATE \textbf{choose} $\vsk \in \vsset$ \textbf{ s.t. } $\forall t \in \vsset$.\ \textsc{Size}($\vsk$) $\leq$ \textsc{Size}($t$)
        \STATE $\vsset \leftarrow \vsset \setminus \{\vsk\}$
        \FOR{$\vsk' \in \textsc{AssignTables}(\vsk,\vec{T}_\text{in})$}
            \FOR{$\vpg \in \textsc{CompleteSketch}(\vsk', T_\text{out}, \mathcal{C})$}
                \IF{$\vpg(\vec{T}_\text{in}) = T_\text{out}$}
                    \RETURN $\vpg$
                \ENDIF
            \ENDFOR
        \ENDFOR
        \STATE $\vsset \leftarrow \vsset \cup \textsc{ExpandSketch}(\vsk)$
    \ENDWHILE
    \RETURN $\bot$
    \end{algorithmic}
    \caption{The top-level synthesis algorithm}
    \label{fig:algo_overview}
\end{figure}

\section{Synthesis Algorithm}
\label{sec:synth_overview}

In this section, we describe the details of our synthesis algorithm. Figure \ref{fig:algo_overview} shows the top-level algorithm. This algorithm takes as input an I/O example $\mathcal{E} = (\vec{T}_\text{in}, T_\text{out})$ and constants $\mathcal{C}$, and returns a synthesized program. The algorithm steps are executed as follows. We first initialize a set of sketches $\vsset$ with 
\added{a singleton having 
}
the sketch $\texttt{Table}(\square)$ or $\texttt{Order}(\texttt{Table}(\square), \square)$, depending on whether $T_\text{out}$ is sorted or not (lines~1-4), and iterate the following operations. In each iteration, we retrieve a sketch $\vsk$ with the minimum size of the sketches in $\vsset$ (lines~6-7). For each sketch $\vsk$, we assign a table name to each `$\square$'  in $\texttt{Table}(\square)$ by calling the function \textsc{AssignTables} (line~8). Then, we complete all of the remaining `$\square$'s in the sketch by calling  \textsc{CompleteSketch} (line~9). When the completion succeeds and a program $\vpg$ is found, we check whether the output table is equal to the evaluation result of $\vpg$ (line~10). If the check succeeds, we return the program $\vpg$ as a result (line~11). Otherwise, we generate additional sketches from the sketch $\vsk$ by calling \textsc{ExpandSketch} (line~12). The operations on lines 5-12 are iterated until a solution is found. Note that this algorithm never stops when there are no solutions in the search space. Therefore, a timeout should be set when it is provided to users in practice.

\added{ 
We show an example of how the top-level algorithm works by using the illustrative example in Figure~\ref{fig:overview}. We start with a singleton with the sketch $s_0 = \texttt{Order}(\texttt{Table}(\square), \square)$ since the output table \texttt{tableOut} is sorted by the \texttt{id} column. First, we try to complete the `$\square$'s in the sketch $s_0$, but cannot find a program consistent with the I/O tables. Then, we expand the sketch $s_0$ by calling \textsc{ExpandSketch} function and obtain new sketches including $s_1 = $\texttt{Order}(\texttt{Select}(\texttt{Table}($\square$), $\square$)), $\square$). We choose the sketch $s_1$ as the next candidate, but we fail to complete the sketch $s_1$ and then expand $s_1$ to obtain new sketches. After repeating the operations, we find the sketch $s$ in Section~\ref{sec:pre_sketch} and successfully complete it. As a result, we can obtain the solution program consistent with the I/O tables, i.e. the program in the center of Figure~\ref{fig:overview}. 
}

Our algorithm returns a program that has the minimum size of the possible solutions since it tries to complete sketches in ascending order of the sketch size. This design is based on Occam's razor, i.e., the hypothesis that the simplest solution is most likely to be correct. This strategy is also helpful for users to understand synthesized queries. In contract, there is prior work that \minor{finds multiple candidates and returns top-k solutions} based on criteria other than the size of programs \cite{scythe, pbe}. 
To make our PBE tool interactive, we return a simple program that satisfies the given specification as quick as possible, rather than returning more sophisticated programs by taking more synthesis time.
\added{ 
In addition, the tolerable waiting time for computer response is known to be about two seconds \cite{waiting_time1, waiting_time2}. Thus, the synthesis time is preferable to be less than two seconds. 
}


\def\Xa{$X_3$}\def\Xb{$X_1$}\def\Xc{$X_2$}\def\Xd{$X_4$}\def\T{$\checkmark$}

\begin{table}[t]
    \centering
    \caption{The restriction of parent-child relations between operators in a sketch. Rows and columns represent parents and children, respectively. The combinations $\checkmark$ are allowed.}
    \label{tab:comb}
    {\small
    \begin{tabular}{lccccccccc}\toprule
          & \rotatebox{90}{\added{\texttt{Order}}} & \rotatebox{90}{\added{\texttt{Distinct}}} & \rotatebox{90}{\added{\texttt{Project}}} & \rotatebox{90}{\added{\texttt{Select}}} & \rotatebox{90}{\added{\texttt{Group}}} & \rotatebox{90}{\added{\texttt{Window}}} & \rotatebox{90}{\added{\texttt{Join}}} & \rotatebox{90}{\added{\texttt{LeftJoin}}}  & \rotatebox{90}{\added{\texttt{Table}}} \\ \midrule
          \texttt{Order}     &\Xa&\T &\T &\T &\T &\T &\T &\T &\T \\
          \texttt{Distinct}  &\Xb&\Xa&\T &\T &\T &\T &\T &\T &\T \\
          \texttt{Project}&\Xb&\Xc&\Xa&\T &\T &\T &\T &\T &\T \\
          \texttt{Select} &\Xb&\Xc&\Xa&\Xa&\T &\T &\T &\T &\T \\
          \texttt{Group}     &\Xb&\Xc&\Xa&\T &\T &\T &\T &\T &\T \\
          \texttt{Window}    &\Xb&\Xc&\Xa&\T &\T &\T &\T &\T &\T \\
          \texttt{Join}      &\Xb&\Xc&\Xa&\Xa&\T &\T &\T &\T &\T \\
          \texttt{LeftJoin}  &\Xb&\Xc&\Xa&\T &\T &\T &\T &\T &\T \\\bottomrule
    \end{tabular}
    }
\end{table}

\subsection{Sketch Generation}
\label{sec:sk_gen}
We describe the details of $\textsc{ExpandSketch}(s)$. This function takes a sketch $\vsk$ and returns  sketches that are created by appending an operator to $\vsk$. Here we refer to the elements other than \texttt{Table} in the right-hand side of $\langle s \rangle$ in Figure \ref{fig:abs_grammar} as \emph{sketch constructors}. This algorithm inserts each sketch constructor above each position of \texttt{Table} in the sketch $\vsk$. First, we find a $\texttt{Table}(\square)$ in $\vsk$ and replace it with a sketch constructor. At this point, the entire sketch contains one or more $\langle s \rangle$ symbols. Then, we replace all of the $\langle s \rangle$s with $\texttt{Table}(\square)$s. 
For example, when the input sketch is $s = \texttt{Project}(\texttt{Table}(\square), \square)$, the result of $\textsc{ExpandSketch}(s)$ includes the following sketches.
\begin{itemize}
    \item $\texttt{Project}(\texttt{Select}(\texttt{Table}(\square), \square), \square)$
    \item $\texttt{Project}(\texttt{Group}(\texttt{Table}(\square), \square, \square), \square)$
    \item $\texttt{Project}(\texttt{Join}(\texttt{Table}(\square), \texttt{Table}(\square), \square), \square)$
\end{itemize}

Here we restrict the combinations of the operators that can appear in a sketch. Table \ref{tab:comb} shows the parent-child relations that are allowed in a syntax tree. The combinations marked as `$\checkmark$' are allowed whereas `$X$' are not. 

We exclude the combinations marked as \Xb\ in Table \ref{tab:comb} because the operators except \texttt{Order} do not preserve the order of records, and therefore the order determined by \texttt{Order} is meaningful only when it is at the top of a sketch. We also exclude \Xc\ because it is considered to be rare in real queries. 
\added{ 
In fact, there exist 1,446 SQL queries having \texttt{DISTINCT} in Spider benchmark~\cite{spider}, which consists of practical SQL queries from various domains, and none of them violate the combination \Xc.
} 
Additionally, we exclude \Xa\ because sketches including the combinations are not in \emph{normal form}. That is, when a program $\vpg$ is obtained from a sketch that is not in normal form, we can always obtain a program equivalent to $\vpg$ from a sketch in normal from. These properties are based on transformation rules in relational algebra. For example, the following rules are known \cite{complete, sql_infer}.
\begin{itemize}
    \item $\texttt{Project}(\texttt{Project}(T, c_1), c_2) \rightarrow \texttt{Project}(T, c_2)$
    \item $\texttt{Select}(\texttt{Select}(T, p_1), p_2) \rightarrow \texttt{Select}(T, p_1 \wedge p_2)$
\end{itemize}
These rules mean that a program that repeats the same operator can always be expressed as a program without repetition, and hence it is sufficient to generate only sketches without repetition in such cases. Besides, the following rules are known.
\begin{itemize}
    \item $\texttt{Select}(\texttt{Project}(T, c), p) \rightarrow \texttt{Project}(\texttt{Select}(T, p), c)$
    \item $\texttt{Join}(\texttt{Project}(T_1, c), T_2, p) \rightarrow \texttt{Project}(\texttt{Join}(T_1, T_2, p), c') $
\end{itemize}
These are rules for moving \texttt{Project} above \texttt{Select} and \texttt{Join}. The same rule is also applicable for \texttt{Group}, \texttt{Window} and \texttt{LeftJoin}. As a consequence, the combinations in Table \ref{tab:comb} allow only sketches with at most one \texttt{Project} and with \texttt{Project} above the operators other than \texttt{Order} and \texttt{Distinct}. Also, the sketch constructors \texttt{Order}, \texttt{Distinct} and \texttt{Project} can appear at the top of the sketch in this order. Importantly, this restriction leads to the efficient completion of the sketch $\texttt{Project}(\vsk', \square)$.
Note that the restrictions \Xb\ and \Xa\ do not decrease the expressiveness of synthesized queries. 

Finding an appropriate program structure from the given I/O specification is essentially difficult in most domains \cite{microsoft-survey}, and our domain of SQL is not an exception. Concretely, the worst-case complexity of the sketch generation algorithm is exponential to the size of the sketch. The reason is that, when creating sketches with the size of $n$, we need to insert each sketch constructor into the sketches with the size of $n-2$ (for \texttt{Window} constructor) or $n-1$ (for the other constructors). The mitigation of the computational cost will help us to find complex program structures. 

\subsection{Sketch Completion}
\label{sec:comp}

\begin{figure}
\def\w{\underline{\hspace{0.2cm}}}
{\small
    \begin{align*}
    \textsc{Propagate}(\varphi, \texttt{Order}) &=\  \varphi \\
    \textsc{Propagate}(\varphi , \texttt{Distinct}) &=\  \begin{cases}
        (M, =_\text{set}) & \text{if } \varphi = (M, =_\text{bag}) \\
        (M, \subseteq_\text{set}) & \text{if } \varphi = (M, \subseteq_\text{bag}) \\
        \ \varphi & \text{otherwise} \\
    \end{cases} \\
    \textsc{Propagate}(\varphi , \texttt{Project}) &=\  \begin{cases}
        (\Mapsto, R) & \text{if } \varphi = ( \Leftrightarrow, R ) \\
        \ \varphi & \text{otherwise} \\
    \end{cases} \\
    \textsc{Propagate}(\varphi , \texttt{Select}) &=\  \begin{cases}
        (M, \subseteq_\text{bag}) & \text{if } \varphi = (M, =_\text{bag}) \\
        (M, \subseteq_\text{set}) & \text{if } \varphi = (M, =_\text{set}) \\
        \ \varphi & \text{otherwise} \\
    \end{cases}
    \end{align*}
    \begin{align*}
    \textsc{Propagate}(\w, \texttt{Group}) &=\  \top &\textsc{Propagate}(\w, \texttt{Window})    &=\  \top \\
    \textsc{Propagate}(\w, \texttt{Join})  &=\  \top &\textsc{Propagate}(\w, \texttt{LeftJoin})  &=\  \top 
    \end{align*}
}
    \caption{Definition of the propagation function}
    \label{fig:prop_function}
\end{figure}%

\begin{figure*}
\small{
\begin{minipage}{.3\textwidth}
    \begin{flushleft}
        \ul{\textsc{CompleteSketch}}$(\vsk, T_\text{out}, \mathcal{C})$\\
        {\small
        \textbf{Input}\ $\vsk$:\ a sketch, $T_\text{out}$, $\mathcal{C}$:\ constants \\
        \textbf{Output}\ a set of programs}
    \end{flushleft}
    \begin{algorithmic}[1]
    \STATE $\varphi_0 \leftarrow (\Leftrightarrow,  =_\text{bag})$
    \RETURN $\textsc{Complete}(\vsk, T_\text{out}, \mathcal{C}, \varphi_0)$
    \end{algorithmic}
    \vspace{0.15cm}
    \begin{flushleft}
        \ul{\textsc{Complete}}$(\texttt{Table}(\text{name}),T_\text{out}, \mathcal{C}, \varphi)$
     \end{flushleft}
    \begin{algorithmic}[1]
    \STATE $\vpg \leftarrow \texttt{Table}(\text{name})$
    \IF{$\varphi(T_\text{out}, \llbracket \vpg \rrbracket) = \top$}
        \RETURN $\{ \vpg \}$
    \ELSE
        \RETURN $\emptyset$        
    \ENDIF
    \end{algorithmic}
    \begin{flushleft}
        \ul{\textsc{Complete}}$(\texttt{Order}(\vsk', \square),T_\text{out}, \mathcal{C}, \varphi)$
     \end{flushleft}
    \begin{algorithmic}[1]
    \STATE \textcolor{darkgray}{// always $\varphi = (\Leftrightarrow,  =_\text{bag})$}
    \STATE $\vpset \leftarrow \emptyset$
    \FOR{$\vpg' \in \textsc{Complete}(\vsk', T_\text{out}, \mathcal{C}, \varphi)$}
        \STATE $\textit{keys} \leftarrow \textsc{SortKeys}(T_\text{out}, \llbracket p' \rrbracket)$
        \STATE $\vpg \leftarrow \texttt{Order}(\vpg', \textit{keys})$
        \STATE \textcolor{darkgray}{// always $\varphi(T_\text{out}, \llbracket \vpg \rrbracket) = \top$}
        \STATE $\vpset \leftarrow \vpset \cup \{ \vpg \}$
    \ENDFOR
    \RETURN $\vpset$
    \end{algorithmic}
    \begin{flushleft}
        \ul{\textsc{Complete}}$(\texttt{Distinct}(\vsk'),T_\text{out}, \mathcal{C}, \varphi)$
     \end{flushleft}
    \begin{algorithmic}[1]
    \STATE $\vpset \leftarrow \emptyset$
    \STATE $\varphi' \leftarrow \textsc{Propagate}(\varphi, \texttt{Distinct})$
    \FOR{$\vpg' \in \textsc{Complete}(\vsk', T_\text{out}, \mathcal{C}, \varphi')$}
        \STATE $\vpg \leftarrow \texttt{Distinct}(\vpg')$
        \IF{$\varphi(T_\text{out}, \llbracket \vpg \rrbracket) = \top$}
            \STATE $\vpset \leftarrow \vpset \cup \{ \vpg \}$
        \ENDIF
    \ENDFOR
    \RETURN $\vpset$
    \end{algorithmic}
\end{minipage}%
\begin{minipage}{.33\textwidth}
    \begin{flushleft}
        \ul{\textsc{Complete}}$(\texttt{Project}(\vsk', \square),T_\text{out}, \mathcal{C}, \varphi)$
     \end{flushleft}
    \begin{algorithmic}[1]
    \STATE \textcolor{darkgray}{// always $\varphi = (\Leftrightarrow, R)$}
    \STATE $\vpset \leftarrow \emptyset$
    \STATE $\varphi' \leftarrow (\Mapsto, R)$ \textcolor{darkgray}{// $= \textsc{Propagate}(\varphi, \texttt{Project})$}
    \FOR{$\vpg' \in \textsc{Complete}(\vsk', T_\text{out}, \mathcal{C}, \varphi')$}
        \FOR{$i \in 1 \dots |T_\text{out}|$}
            \STATE $R_i \leftarrow \{ i' \mid R(\llbracket \vpg' \rrbracket[i'], T_\text{out}[i])\}$
        \ENDFOR
        \FOR{$\textit{cols} \in R_1 \times \dots \times R_{|T_\text{out}|}$}
            \STATE $\vpg \leftarrow \texttt{Project}(\vpg', \textit{cols})$
            \IF{$\varphi(T_\text{out}, \llbracket \vpg \rrbracket) = \top$}
                \STATE $\vpset \leftarrow \vpset \cup \{ \vpg \}$
            \ENDIF
        \ENDFOR
    \ENDFOR
    \RETURN $\vpset$
    \end{algorithmic}
    \begin{flushleft}
        \ul{\textsc{Complete}}$(\texttt{Select}(\vsk', \square),T_\text{out}, \mathcal{C}, \varphi)$
     \end{flushleft}
    \begin{algorithmic}[1]
    \STATE $\vpset \leftarrow \emptyset$
    \STATE $\varphi' \leftarrow \textsc{Propagate}(\varphi, \texttt{Select})$
    \FOR{$\vpg' \in \textsc{Complete}(\vsk', T_\text{out}, \mathcal{C}, \varphi')$}
        \FOR{$\textit{cond} \in \textsc{Conds}(\llbracket \vpg' \rrbracket, \mathcal{C})$}
            \STATE $\vpg \leftarrow \texttt{Select}(\vpg', \textit{cond})$
            \IF{$\varphi(T_\text{out}, \llbracket \vpg \rrbracket) = \top$}
                \STATE $\vpset \leftarrow \vpset \cup \{ \vpg \}$
            \ENDIF
        \ENDFOR
    \ENDFOR
    \RETURN $\vpset$
    \end{algorithmic}
    \begin{flushleft}
        \ul{\textsc{Complete}}$(\texttt{Window}(\vsk', \square),T_\text{out}, \mathcal{C}, \varphi)$
     \end{flushleft}
    \begin{algorithmic}[1]
    \STATE $\vpset \leftarrow \emptyset$
    \FOR{$\vpg' \in \textsc{Complete}(\vsk', T_\text{out}, \mathcal{C}, \top)$}
        \STATE $\textit{wins} \leftarrow \textsc{Wins}(\llbracket \vpg' \rrbracket)$
        \STATE $\vpg \leftarrow \texttt{Window}(\vpg', \textit{wins})$
        \IF{$\varphi(T_\text{out}, \llbracket \vpg \rrbracket) = \top$}
            \STATE $\vpset \leftarrow \vpset \cup \{ \vpg \}$
        \ENDIF
    \ENDFOR
    \RETURN $\vpset$
    \end{algorithmic}
\end{minipage}%
\begin{minipage}{.36\textwidth}
    \begin{flushleft}
        \ul{\textsc{Complete}}$(\texttt{Group}(\vsk', \square, \square),T_\text{out}, \mathcal{C}, \varphi)$
     \end{flushleft}
    \begin{algorithmic}[1]
    \STATE $\vpset \leftarrow \emptyset$
    \FOR{$\vpg' \in \textsc{Complete}(\vsk', T_\text{out}, \mathcal{C}, \top)$}
        \STATE $\textit{aggs} \leftarrow \textsc{Aggs}(\llbracket \vpg' \rrbracket)$
        \FOR{$\textit{cols}\ \textbf{ s.t. }\ \textit{cols} \subseteq \textsc{Cols}(\llbracket \vpg' \rrbracket) \wedge |\textit{cols}| \leq 2$}
            \STATE $\vpg \leftarrow \texttt{Group}(\vpg', \textit{cols}, \textit{aggs})$
            \IF{$\varphi(T_\text{out}, \llbracket \vpg \rrbracket) = \top$}
                \STATE $\vpset \leftarrow \vpset \cup \{ \vpg \}$
            \ENDIF
        \ENDFOR
    \ENDFOR
    \RETURN $\vpset$
    \end{algorithmic}
    \begin{flushleft}
        \ul{\textsc{Complete}}$(\texttt{Join}(\vsk'_1, \vsk'_2, \square),T_\text{out}, \mathcal{C}, \varphi)$
     \end{flushleft}
    \begin{algorithmic}[1]
    \STATE $\vpset \leftarrow \emptyset$
    \STATE{$\vpset'_1 \leftarrow \textsc{Complete}(\vsk'_1, T_\text{out}, \mathcal{C}, \top)$}
    \STATE{$\vpset'_2 \leftarrow \textsc{Complete}(\vsk'_2, T_\text{out}, \mathcal{C}, \top)$}
    \FOR{$(\vpg'_1, \vpg'_2) \in \vpset'_1 \times \vpset'_2$}
        \FOR{$\textit{pairs} \in \textsc{Pairs}(\llbracket \vpg'_1 \rrbracket, \llbracket \vpg'_2 \rrbracket)$}
            \STATE $\vpg \leftarrow \texttt{Join}(\vpg'_1, \vpg'_2, \textit{pairs})$
            \IF{$\varphi(T_\text{out}, \llbracket \vpg \rrbracket) = \top$}
                \STATE $\vpset \leftarrow \vpset \cup \{ \vpg \}$
            \ENDIF
        \ENDFOR
    \ENDFOR
    \RETURN $\vpset$
    \end{algorithmic}
    \begin{flushleft}
        \ul{\textsc{Complete}}$(\texttt{LeftJoin}(\vsk'_1, \vsk'_2, \square),T_\text{out}, \mathcal{C}, \varphi)$
     \end{flushleft}
    \begin{algorithmic}[1]
    \STATE $\vpset \leftarrow \emptyset$
    \STATE{$\vpset'_1 \leftarrow \textsc{Complete}(\vsk'_1, T_\text{out}, \mathcal{C}, \top)$}
    \STATE{$\vpset'_2 \leftarrow \textsc{Complete}(\vsk'_2, T_\text{out}, \mathcal{C}, \top)$}
    \FOR{$(\vpg'_1, \vpg'_2) \in \vpset'_1 \times \vpset'_2$}
        \FOR{$\textit{pair} \in \textsc{Pair}(\llbracket \vpg'_1 \rrbracket, \llbracket \vpg'_2 \rrbracket)$}
            \STATE $p \leftarrow \texttt{LeftJoin}(\vpg'_1, \vpg'_2, \textit{pair})$
            \IF{$\varphi(T_\text{out}, \llbracket \vpg \rrbracket) = \top$}
                \STATE $\vpset \leftarrow \vpset \cup \{ \vpg \}$
            \ENDIF
        \ENDFOR
    \ENDFOR
    \RETURN $\vpset$
    \end{algorithmic}
    \end{minipage}
}
    \caption{Algorithms for sketch completion}
    \label{fig:algo_complete_sketch}
\end{figure*}

First, we describe the function $\textsc{AssignTables}(\vsk,\vec{T}_\text{in})$ used in the top-level algorithm in Figure \ref{fig:algo_overview}. This function takes as input a sketch $\vsk$ and the input tables $\vec{T}_\text{in}$, and returns sketches with the tables' name filled. Specifically, it assigns a table name in $\vec{T}_\text{in}$ to each $\texttt{Table}(\square)$ in the sketch $\vsk$. 
Here all of the input tables must be used to complete the sketch. This limitation is based on the assumption that the user does not give input tables that are not used in a resulting query, and it is effective for excluding sketches that do not meet the user's intention.
For example, when a sketch $s = \texttt{LeftJoin}(\texttt{Table}(\square), \texttt{Table}(\square), \square)$ and $\vec{T}_\text{in} = \{ T_1, T_2 \}$, the result of $\textsc{AssignTables}(\vsk,\vec{T}_\text{in})$ includes the following sketches.
\begin{itemize}
    \item $\texttt{LeftJoin}(\texttt{Table}(T_1), \texttt{Table}(T_2), \square)$
    \item $\texttt{LeftJoin}(\texttt{Table}(T_2), \texttt{Table}(T_1), \square)$
\end{itemize}
Note that the sketch $\texttt{LeftJoin}(\texttt{Table}(T_1), \texttt{Table}(T_1), \square)$ is not included since this does not use the input table $T_2$.

Next, we explain the function $\textsc{CompleteSketch}(\vsk, T_\text{out}, \mathcal{C})$. This function takes as input a sketch $\vsk$, the output table $T_\text{out}$ and constants $\mathcal{C}$, and returns a set of programs. It fills `$\square$'s in the sketch $\vsk$ by using constants in $\mathcal{C}$, and thus constructs programs whose evaluation result can be $T_\text{out}$. Figure \ref{fig:algo_complete_sketch} shows the algorithms for sketch completion. The function  \textsc{CompleteSketch}$(s, T_\text{out}, \mathcal{C})$ is the entry point, and it invokes auxiliary functions named \textsc{Complete} for recursively completing a sketch by propagating a constraint $\varphi$. 
\added{ 
The completion algorithms are similar to those used in \morpheus~\cite{morpheus} with two main differences. First, we prune the search space based on table inclusion relation $\varphi$ while \morpheus\ relies on the metadata of tables and employs an SMT solver. Second, the completion algorithm of projection sketches is novel and efficient.
}

Before describing the details of each function in Figure \ref{fig:algo_complete_sketch}, we start with the notations that are commonly used in these functions.
We use $\llbracket \vpg \rrbracket$ to refer to the table that can be gained by evaluating a program $\vpg$. 
The function $\textsc{Cols}(T)$ returns the columns in table $T$.
$\textsc{Propagate}(\varphi, \textit{op})$ takes a constraint $\varphi$ and an operator type $\textit{op}$, and returns a constraint $\varphi'$. Figure \ref{fig:prop_function} shows the definition of $\textsc{Propagate}$. It calculates a precondition $\varphi'$ from the postcondition $\varphi$ of the sketch completion. Accurately, suppose we have a sketch $\vsk$ and its child $\vsk'$, and the programs $\vpg$ and $\vpg'$ are obtained by completing $\vsk$ and $\vsk'$, respectively. When the sketch $\vsk$ has the operator $\textit{op}$, $\textsc{Propagate}(\varphi, \textit{op})$ returns a constraint $\varphi'$ such that  $\varphi'(T_\text{out}, \llbracket \vpg' \rrbracket) = \top$, which is a necessary condition for $\varphi(T_\text{out}, \llbracket \vpg \rrbracket) = \top$ to hold. For example, suppose a sketch $\vsk = \texttt{Select}(\vsk', \square)$ is to be completed, and the postcondition is $\varphi = (\Mapsto, =_\text{set})$. This means that $\llbracket \vpg \rrbracket$ needs to have a column equivalent to each column in $T_\text{out}$ as set ($=_\text{set}$). Then, $\textsc{Propagate}(\varphi, \texttt{Select})$ returns $\varphi' = (\Mapsto, \subseteq_\text{set})$ as a precondition for the completion of $\vsk$. The reason is that \texttt{Select} operator does not yield any new values, and therefore $\llbracket \vpg' \rrbracket$ needs to have a column that is superset  ($\subseteq_\text{set}$) of each column in $T_\text{out}$.

We explain each part of the algorithm in Figure \ref{fig:algo_complete_sketch}.
For the completion of $\texttt{Table}(\text{name})$, there are no `$\square$'s to be filled, and  only pruning is performed using the propagated constraint $\varphi$. Here the evaluation result $\llbracket \vpg \rrbracket$ is the same as one of the input tables. Therefore, the pruning essentially compares an input table $T_\text{in}$ and the output table $T_\text{out}$. This is the first pruning that we perform before filling `$\square$'s in a sketch, and here we validate whether the entire sketch can satisfy the I/O tables. For example, if $T_\text{out}$ has a cell with a value $v$ that does not exist in $T_\text{in}$, we can discard the sketch $\texttt{Project}(\texttt{Select}(\texttt{Table}(T_1), \square), \square)$ before filling the remaining `$\square$'s. The reason is that the propagated constraint $\varphi = (\Mapsto, \subseteq_\text{bag})$ can detect the fact that no matter how the remaining `$\square$'s are filled, the value $v$ will not be yielded by \texttt{Select} or \texttt{Project}. This pruning is similar to the validation of a sketch in \morpheus\ \cite{morpheus}. While it uses an SMT solver to validate a sketch,  our algorithm uses a propagated constraint $\varphi$, which considers an inclusion relation between the output and intermediate tables.

For the completion of $\texttt{Order}(s', \square)$, we first complete the child $s'$ and then fill the target sketch with sort keys. The function \textsc{SortKeys} infers sort keys from the columns in $T_\text{out}$. Accurately, we find a column $c_1$ sorted in ascending or descending order from the columns in $T_\text{out}$. If the values in $c_1$ are duplicated, we group the records by the key of $c_1$. Then, we again find sort keys for each group. If $c_2$ is a valid key for all the groups, we obtain a composite key $[c_1, c_2]$. By finding sort keys in the same manner, we obtain a composite key $[c_1, c_2, \dots, c_n]$. Finally, we convert it to the corresponding columns in $\llbracket p' \rrbracket$. Note that pruning with the constraint $\varphi$ is not needed here because the child's constraint $\varphi(T_\text{out}, \llbracket p' \rrbracket) = \top$ holds and hence the predicate $\varphi(T_\text{out}, \llbracket p \rrbracket) = \top$ is always true. 

For the completion of $\texttt{Distinct}(s')$,  we complete the child sketch and discard invalid programs since there are no `$\square$'s to be filled,.

For the completion of $\texttt{Project}(s', \square)$, we fill the sketch with projected columns. This  is one of the most distinctive parts of our algorithm. Importantly, owing to the restriction of sketch structures in Table \ref{tab:comb}, the constraint $\varphi$ always has the form of $\varphi = (\Leftrightarrow, R)$ when this function is invoked. 
First, we obtain the program $\vpg'$ by completing the child of the sketch. Then, for each $i$-th column in $T_\text{out}$, we calculate the set $R_i$ of indices $i'$ such that the $i'$-th column in $\llbracket p' \rrbracket$ corresponds to the $i$-th column in $T_\text{out}$ by relation $R$. Note that every $R_i$ cannot be empty because the child's constraint $\varphi' = (\Mapsto, R) \wedge \varphi'(T_\text{out}, \llbracket p' \rrbracket) = \top$ holds. Finally, we enumerate the elements in the cartesian product of $R_1,\dots,R_{|T_\text{out}|}$ as the candidates for projected columns.
The bottlenecks of this part are (1) the calculation of each $R_i$ and (2) the calculation of the cartesian product of $R_1, \dots, R_{|T_\text{out}|}$. The computational complexity of (1) is $O(\| T_\text{out} \| \| \llbracket p' \rrbracket \|)$, where $\| T \|$ is the number of the cells in table $T$. The worst-case complexity of (2) is exponential, but the computation does not cause a combinatorial explosion in most cases because the size of each $R_i$ is generally small. 

In contrast, prior work  \cite{scythe, morpheus} does not expect that the columns in $\llbracket p' \rrbracket$ correspond to those in $T_\text{out}$ in their order (i.e., `$\Leftrightarrow$' in our work) when \texttt{Project} sketch is being completed. Therefore, it needs to enumerate all the permutations of the columns in $\llbracket p' \rrbracket$ as in Figure \ref{fig:algo_prior_proj}. The computational cost grows exponentially as the number of the columns in $\llbracket p' \rrbracket$ increases, which leads to a scalability issue in the overall algorithm.

For the completion of $\texttt{Select}(s', \square)$, we fill the sketch with predicates. The function $\textsc{Conds}(T, \mathcal{C})$ returns the predicates using columns in table $T$ and constants in $\mathcal{C}$. These predicates conform to the rule $\langle \textit{pred} \rangle$ in the grammar of Figure \ref{fig:grammar}, and they have consistent types between columns and constants. To enumerate the predicates efficiently, the information about remaining rows is encoded into a \textit{bit array}, which is introduced in \scythe\ \cite{scythe}. That is, we calculate a bit array $b$ for each predicate, where $b[i] = 1$ if the $i$-th row in $T$ remains, and $b[i] = 0$ otherwise. Specifically, we enumerate predicate candidates in the order of $\langle \textit{prim} \rangle$, $\langle \textit{clause} \rangle$ and $\langle \textit{pred} \rangle$. A predicate for $\langle \textit{clause} \rangle$ is a disjunction of $\langle \textit{prim} \rangle$s, and a $\langle \textit{pred} \rangle$ is a conjunction of $\langle \textit{clause} \rangle$s. When calculating the remaining rows as a result of these predicates, we perform the operations $\vee$ and $\wedge$ over bit arrays instead of executing compound predicates over an instantiated table. Since we try to find a single program that satisfies the I/O tables, we discard predicates that have the same bit array as previously enumerated ones.

For the completion of $\texttt{Window}(s', \square)$, we fill the sketch with the columns of window functions. The function $\textsc{Win}(T)$ enumerates the columns in the form of  $\langle win \rangle$ in the grammar of Figure \ref{fig:grammar}, i.e., a window function, a target column, partitioning keys and a sort key. The number of these combinations may be large in some cases, and the table obtained by $\llbracket p \rrbracket$ possibly has a large number of columns. However, this does not significantly affect the overall performance since our algorithm is resistant to the increase in column size. This strategy enables our algorithm to do without user hints as to which functions should be used. Of course, due to this strategy, synthesized programs may contain useless columns that do not affect the behavior. Hence, we eliminate such columns after we have found a correct program.

For the completion of $\texttt{Group}(s', \square, \square)$, we fill the sketch with grouping keys and aggregation columns. Similar to the completion of $\texttt{Window}$ sketch, the function $\textsc{Aggs}(\llbracket p' \rrbracket)$ enumerates the combinations of the columns in $\llbracket p' \rrbracket$ and the aggregation functions. We then enumerate grouping keys, the size of which is less than or equal to two. We discuss this limitation in Section \ref{sec:adapt}.

For the completion of $\texttt{Join}(s'_1, s'_2, \square)$ and $\texttt{LeftJoin}(s'_1, s'_2, \square)$, we fill the sketch with key pair(s). These sketches have two children $s'_1$ and $s'_2$, and we begin with completing them to obtain the sets $P'_1$ and $P'_2$ of programs. Then, we fill join predicates for each pair of the programs $p'_1 \in P'_1$ and $p'_2 \in P'_2$. The function $\textsc{Pairs}(T_1, T_2)$  uses columns in $T_1$ and $T_2$ to create the predicates in the form of $\langle pairs \rangle$ in Figure \ref{fig:grammar}. In particular, it first enumerates the predicates for $\langle pair \rangle$, and then combines them using $\wedge$ to generate the conjunctive predicates for $\langle pairs \rangle$. Similarly, $\textsc{Pair}(T_1, T_2)$ returns a set of single pairs in $T_1$ and $T_2$ for the completion of \texttt{LeftJoin} sketches. In these processes, we use bit arrays to efficiently enumerate the predicates as in the completion of $\texttt{Select}$ sketches. In addition to calculating remaining rows efficiently, we reduce the cost for instantiating the cross product of two tables, as in \scythe.

\begin{figure}[t]
    \begin{flushleft}
        \ul{\textsc{Complete}}$(\texttt{Project}(\vsk', \square),T_\text{out}, \mathcal{C}, \varphi)$
     \end{flushleft}
    \begin{algorithmic}[1]
    \STATE $\vpset \leftarrow \emptyset$
    \FOR{$\vpg' \in \textsc{Complete}(\vsk', T_\text{out}, \mathcal{C}, \varphi)$}
        \FOR{$\textit{cols} \subseteq \textsc{Cols}(\llbracket p' \rrbracket)$}
            \FOR{$\textit{cols}' \in \textsc{Permute}(cols)$}
                \STATE $\vpg \leftarrow \texttt{Project}(\vpg', \textit{cols}')$
                \IF{$\varphi(T_\text{out}, \llbracket \vpg \rrbracket) = \top$}
                    \STATE $\vpset \leftarrow \vpset \cup \{ \vpg \}$
                \ENDIF
            \ENDFOR
        \ENDFOR
    \ENDFOR
    \RETURN $\vpset$
    \end{algorithmic}
    \caption{The completion of \texttt{Project} sketch in prior work}
    \label{fig:algo_prior_proj}
\end{figure}

\subsection{Adaptability to Grammar Extensions}
\label{sec:adapt}
Our algorithm can easily support additional syntax features for predicates and functions because it does not depend on the concrete semantics of them. For example, we can easily support the aggregation function \texttt{STDEV}, which calculates the statistical standard deviation for each group, by adding it to the list of aggregation functions. Also, supporting \texttt{LIKE} in predicates is straightforward if the user provides pattern strings used in \texttt{LIKE} predicates as part of constants $\mathcal{C}$.

In contrast, it is difficult for our algorithm to naively support the \texttt{UNION} clause, which is an operator to combine the records in two tables. Because the positions of \texttt{Union} and \texttt{Project} cannot be safely interchanged in sketch structures, we can no longer fix the position of \texttt{Project} above other operators as the combinations in Table \ref{tab:comb}. Hence, filling $\texttt{Project}(s', \square)$ with projected columns leads to a significant increase in computational cost as in prior work. However, in general, the \texttt{UNION} clause is used just for combining the results from multiple subqueries. We believe that the user does not have difficulty combining such queries by using the \texttt{UNION} keyword as long as our method synthesizes each query.
It is also difficult to naively support queries with a large number of grouping keys. The reason is that a slight change in key selection can have a significant effect on yielded values, and we cannot find desired grouping keys without executing queries. Thus, allowing an arbitrary number of grouping keys results in a combinatorial explosion during the completion of \texttt{Group} and \texttt{Window} sketches. This is the reason for which we limit the size of grouping keys to two or less in Figure \ref{fig:grammar}. 

\section{Implementation}
\label{sec:impl}
We implemented this algorithm in Java as a tool \tool\ and optimized it in several points to reduce the search space on sketch structures. First, we do not enumerate the sketches that are ``symmetric'' with the previously explored ones. Accurately, we do not distinguish the sketches in the form of $\texttt{Join}(s_1, s_2, \square)$ and $\texttt{Join}(s_2, s_1, \square)$ since the programs derived from them are semantically equivalent. 
Second, we do not enumerate the sketches that do not contain \texttt{Select} operator when the constants $\mathcal{C}$ are not empty. 

We provide a user interface that is inspired by the concept of \emph{live programming}. Live programming is an interactive programming environment, where the user can get real-time feedback on the behavior of a program each time s/he updates the code \cite{live_prog}. Similarly, we implemented an interactive user interface for PBE, where the user can get real-time feedback on the synthesized program each time s/he updates the I/O example. This interface enables the user to start with a simple example and progressively create more complex examples, as long as the synthesis time is reasonably short.

\newcommand{~}{\phantom{0}}
\begin{table*}[t]
    \centering
    \caption{The number of benchmarks solved by different algorithms. The ``No.'' column represents the number of queries. The ``\#Col'' and ``\#Cell'' columns represent the average number of columns and cells, respectively.}
    \label{tab:result_benchmark}
    \begin{tabular}{cccc|cccc}\hline
    Benchmark      & No.  &\#Col &\#Cell & \tool & \added{\tool$_\text{5}$} & \scythe & \textsc{BaseLine} \\ \hline
    \texttt{ase13}  & ~28   & ~3.2 & ~50.5 & \textbf{~25 (89\%)} & ~21 (75\%) & ~15 (54\%)  & ~17 (60\%) \\
    \texttt{so-top} & ~57   & ~3.0 & ~17.2 & \textbf{~42 (74\%)} &  \textbf{~42 (74\%)} & ~40 (70\%)  & ~30 (53\%) \\
    \added{\texttt{so-dev}} & ~57  &  ~4.9 & ~26.1 & \textbf{~47 (82\%)} & ~46 (81\%) & ~46 (81\%)   & ~38 (67\%)  \\
    \added{\texttt{so-rec}} & ~51   & ~5.5 & ~33.7 & ~20 (39\%)  & ~17 (33\%) & \textbf{~27 (53\%)} & ~13 (25\%) \\
    \texttt{kaggle} & ~33   & 12.1 & 164.5 & \textbf{~19 (58\%)} & ~18 (55\%) & ~~0 (~0\%)  & ~~1 (~3\%) \\ \hline
     total          &226   & ~~-~ &  ~~-~ & \textbf{153 (68\%)} & 144 (63\%) & 128 (57\%) &  ~99 (44\%)\\ \hline
    \end{tabular}
\end{table*}

\section{Evaluation}
\label{sec:eval}
To evaluate \tool, we perform experiments to answer the following research questions.
\begin{itemize}
    \item[\textbf{RQ1:}] Is \tool\ more effective for synthesizing complex SQL queries than prior methods?
    \item[\textbf{RQ2:}] \minor{Do our algorithm improvements perform better than the naive algorithm?}
    \item[\textbf{RQ3:}] What kind of queries does \tool\ fail to synthesize?
    \item[\textbf{RQ4:}] \added{Can \tool\ handle large I/O tables better than prior methods?}
\end{itemize}
To answer the questions, we synthesize a wide variety of SQL queries by using \tool\ and other methods. The experiments are conducted on a machine with 2.20 GHz Intel Xeon CPU and 8 GB of physical memory running the Windows 10 OS.
\subsection{Benchmarks}
\label{sec:eval_benchmarks}
To investigate the effectiveness of \tool\ on practical SQL queries, we collected 226 queries in total from the following benchmarks.
\\
\textbf{\texttt{ase13}:}\ These 28 queries were extracted from a textbook for database systems. The I/O tables were introduced for the evaluation of SQLSynthesizer \cite{sql-synthesizer} and also used for \scythe\  study \cite{scythe}. These queries range from basic ones that non-experts may have difficulty writing, to complex ones that combine standard SQL features. 
\\
\added{
\textbf{\texttt{so-top}, \texttt{so-dev}, \texttt{so-rec}:}\ The three benchmarks have 57, 57 and 51 queries, respectively. They were extracted from Stack~Overflow. The I/O tables were introduced in the evaluation of \scythe. The posts in these benchmarks are questions about SQL programming. The benchmark \texttt{so-top} consists of posts with more than 30 votes, and it represents common and practical issues that developers are faced with. The benchmarks \texttt{so-dev} and \texttt{so-rec} are questions that were posted during the development of \scythe. These posts have considerably fewer votes than \texttt{so-top}, and the intention of the posts tends to be ambiguous. Also, some posts were modified after \scythe\ study. Since \scythe\ study does not publish queries marked as solutions, we determined the solutions based on the latest posts and published them in our Github repository.
}
\added{ 
Note that the median first response time for the posts took 12 minutes, and the median acceptance time took 81 minutes. 
}
\\
\textbf{\texttt{kaggle}:}\ These 33 queries were extracted from SQL tutorials in Kaggle, which is an online community for data science. In particular, we collected queries from tutorials and exercises in 
``Intro to SQL''~\footnote{\url{https://www.kaggle.com/learn/intro-to-sql}} and 
``Advanced SQL''~\footnote{\url{https://www.kaggle.com/learn/advanced-sql}}. 
These include various queries that are common in the context of data analysis. Importantly, the queries handle real tables that are published as datasets in Kaggle. This benchmark is mainly used for evaluating the effectiveness of \minor{the various PBE methods} on I/O tables having full-scale schemas. 
\subsection{Compared Systems}
\minor{
We compare \tool\ with the following methods. 
}
\\
\textbf{\scythe}\ is a state-of-the-art tool that synthesizes complex SQL queries form I/O tables \cite{scythe}. In particular, it supports the synthesis of nested query along with grouping and aggregation by enumerating \textit{abstract queries} in a bottom-up manner. There are several differences in specifications between \tool\ and \scythe. For instance, while \tool\ finds a single solution as a result, \scythe\ finds five solutions that have high scores based on a ranking heuristic. Also, \scythe\ requires hints about which aggregation functions should be used in the resulting query, in addition to constants. 
\\
\added{
\textbf{\tool$_\text{5}$}\ is a top-5 implementation of \tool\ that returns five solutions based on a ranking heuristic. This algorithm is used for a fair comparison with \scythe\ since it also returns five solutions as a result. The ranking heuristic is similar to \scythe. Namely, it considers the simplicity of synthesized queries and the coverage of given constants, and it finds five best solutions among programs of the same size. This algorithm is also used for investigating the effectiveness of \tool's  strategy that finds a single solution. 
}
\\
\textbf{\textsc{BaseLine}}\ is a baseline algorithm that has the same search space as \tool. Specifically, \textsc{BaseLine} synthesizes SQL queries in our DSL in Figure \ref{fig:grammar}, and the overall algorithm is almost the same as \tool. 
\minor{
The difference is, while completing the sketch $\texttt{Project}(s', \square)$, it uses a brute-force search algorithm (Figure \ref{fig:algo_prior_proj}), which is highly inefficient way for finding the projected columns, as opposed to our algorithm (Figure \ref{fig:algo_complete_sketch}). 
}
Also, \textsc{BaseLine} requires hints about which aggregation and window functions should be used. Without the hints, the computationan becomes intractable since the completion of \texttt{Project} sketches needs to calculate the permutations of a large number of columns, which we have appended in the completion of \texttt{Group} and \texttt{Window} sketches. 

\subsection{Evaluation Process}
We performed the synthesis of the benchmark queries by using \tool, \scythe, \added{\toolf}\ and \textsc{BaseLine}.  
For the benchmarks \added{other than \texttt{kaggle}}, we reused the I/O tables that were created in \scythe\ study \cite{scythe} \added{for a fair comparison between these methods. 
It is important to use similar tables for each method since the performance can depend on the size of tables.} The columns related to the query are extracted in advance, and hence the schemas of the I/O tables are not necessarily the same as original ones. For the \texttt{kaggle} benchmark, we created I/O tables for each query. Specifically, the tables were created by a third person, who is familiar with SQL but was not involved in the development of our synthesis algorithm. Here the schemas of I/O tables are the same as those of the original tables in datasets. Thus, the tables in the \texttt{kaggle} benchmark have an average of 12.1 columns, which is larger than the average of \added{4.3} columns \minor{in the other four benchmarks}. In general, using original schemas in I/O examples is preferable for users since there is no need to decide which columns to be used \minor{while creating tables}. 
We provided each algorithm with hints about the constants used in predicates, and we also provided \scythe\ and \textsc{BaseLine} with hints about aggregation and window functions. 
\scythe\ required a total of 110 hints about aggregate functions, some of which are used in the same query. 

The correctness of the synthesized queries was verified by at least two people familiar with SQL. When the synthesized query did not have the same semantics as the solution, we updated the I/O tables to make the intention clearer mainly by adding rows. We continued to update the tables and hints until \minor{a solution} was found or a timeout occurred. 
\added{When a solution was found, we reported the synthesis time of the last execution. Since the synthesis time increases monotonically during the iterations, the time required for the last execution is a good indicator for the total time for multiple iterations.} 
For the evaluation of \scythe\ \added{and \toolf}, we checked if a desired query was included in the five queries returned as a result. We did not check the correctness of the \scythe's results for the benchmarks \texttt{ase13} and \texttt{so-top} because the correctness had been carefully verified in \scythe\ study. 
Since we focus on an interactive usage of a PBE tool as mentioned in Section \ref{sec:impl}, we did not count the number of the updates performed during the synthesis process. 
For the same reason, we performed each synthesis with a time limit of 100 seconds, which is shorter than that in \scythe\ study. 

\begin{figure}[t]
    \centering
    \includegraphics[width=0.98\linewidth]{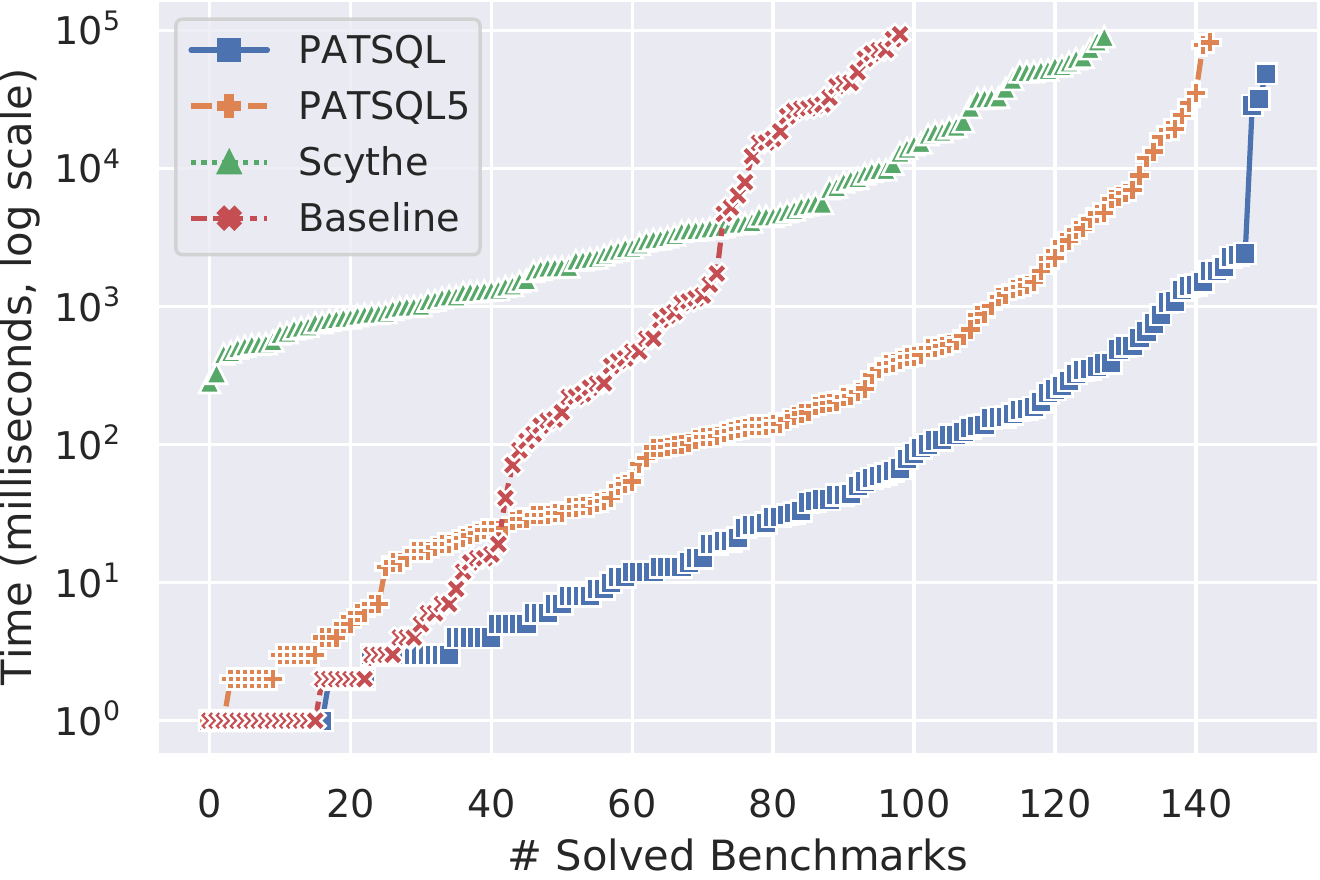}
    \caption{The synthesis time for benchmarks.}
    \label{fig:time_benchmark}
\end{figure}%

\subsection{Comparison to Prior Work}
We first compare the number of benchmarks solved \tool\ and \scythe. 
For the benchmarks other than \texttt{kaggle}, \tool\ solved \added{134} benchmarks while \scythe\ solved \added{128} (see Table \ref{tab:result_benchmark}). In particular, \tool\ succeeded in synthesizing \added{20} queries from I/O tables with a larger number of cells that \scythe\ failed to solve due to scalability issues. Additionally, \tool\ succeeded in synthesizing a query with the window function \added{\texttt{RANK()}}, which \scythe\ does not support. In contrast, \tool\ failed to synthesize \added{15} queries that \scythe\ succeeded in. 
\added{Six cases} were due to non equi-join, which uses inequality operators such as `\texttt{<}' in a \texttt{JOIN} condition, and \added{four cases} were due to the \texttt{UNION} clause. \added{The other causes include unsupported predicates and the order in which sketches were generated.}
For the \texttt{kaggle} benchmark, \tool\ solved 19 benchmarks while \scythe\ did not solve any of them. Since the benchmark handles I/O tables having full-scale schema, \scythe\ caused combinatorial explosions in the enumeration of abstract queries. In contrast, \tool\ succeeded in synthesizing such queries \minor{due to improvements in our algorithm to} support I/O tables having larger schemas. 
In summary, the expressiveness of \tool\ is comparable to that of \scythe, and  \tool\ can synthesize queries from I/O tables having larger schemas that \minor{\scythe\ fails to deal with}. \minor{Because of its highly scalable nature, \tool\ was able to synthesize more queries within a time limit.}
\added{ 
In addition, we count the number of the rows in I/O tables for \tool\ and \scythe\ to find a solution. We found that the rows required for \tool\ is on average 0.56 more than those required for \scythe. This means \tool\ does not impose a large overhead on the user. 
}
\added{
When we compare \toolf\ to \scythe, both of which return top five solutions from the same size programs, \toolf\ still outperforms \scythe\ in the number of solved benchmarks (Table~2). 
}

Next, we compare \tool\ to \scythe\ in terms of the execution time. 
\tool\ synthesized \added{102} of solved benchmarks within 0.1 seconds and \added{136} within a second, while \scythe\ synthesized \added{28} within a second and \added{100} within ten seconds (Figure \ref{fig:time_benchmark}). The faster execution time of \tool\ is useful for interactive scenarios where the user modifies the I/O example step by step. 
\added{
Also, 96\% of the response time taken by \tool\ was shorter than two seconds, i.e. the tolerance waiting time for web response (Section \ref{sec:synth_overview}), and then much shorter than 12 minutes, i.e. the median time for the first response in Stack~Overflow (Section \ref{sec:eval_benchmarks}). Thus, the contributions of \tool\ are shown to be valuable for the practical application of PBE methods.
}
\added{
When we compare \toolf\ to \scythe\ in terms of the execution time, \toolf\ still outperforms \scythe\ (Figure~\ref{fig:time_benchmark}). This shows that \tool's efficiency is significant, even taking into account the difference in the number of solutions returned.  
}

\added{
We compare the performance of \tool\ and \toolf. The solved benchmarks by \toolf\ was slightly fewer than that of \tool\ (Table~\ref{tab:result_benchmark}). The execution time of \toolf\ was about 10 times slower than \tool\ (Figure~\ref{fig:time_benchmark}). This result is as expected since \toolf\ has a larger search space than \tool\ to find multiple candidate programs. We found that 90\% of the I/O examples required for \toolf\ were the same as those required for \tool. In other words, an interface that returns a single solution does not impose significant additional overhead on the user. Thus, we can say that the search priorities based on program simplicity in \tool\ is effective as well as the ranking function in \toolf. In general, a top-k algorithm only works well when it can find the desired solution as one of the candidates and rank it within top k among the candidates. Since there can be an exponential number of candidates in the search space, top-k algorithms tend to require additional examples to resolve ambiguities, as in the algorithm that finds a single solution. Hence, the search strategy that finds a single solution as quickly as possible would be suitable in many practical applications. 
}

For \textbf{RQ1}, the experimental result suggests that \tool\ outperforms a state-of-the-art algorithm \scythe\ in terms of the execution time and the scalability of I/O tables while maintaining the expressiveness of synthesized queries. 
\added{
We also show the effectiveness of \tool's strategy that finds a single solution.
}

\begin{table}[t]
    \centering
    \caption{The causes of \tool's failure. The `f-elem' row means unsupported syntax features, and `f-struct' means complex structures of desired queries that can be represented in DSL but cannot be found within a time limit.}
    \label{tab:failed_cases}
    \begin{tabular}{l|ccccc|c}\hline
                      & \texttt{ase13} & \texttt{so-top} & \texttt{so-dev} & \texttt{so-rec} & \texttt{kaggle} & total\\ \hline
         f-elem~ & 2 & 14 & ~5 & 29  & 10 & 60\\
        f-struct & 1 & ~1 & ~5 & ~2  & ~4 & 13\\ \hline
          total  & 3 & 15 & 10 & 31  & 14 & 73\\ \hline
    \end{tabular}
\end{table}

\begin{figure*}[t]
    \centering
    \includegraphics[width=.33\linewidth]{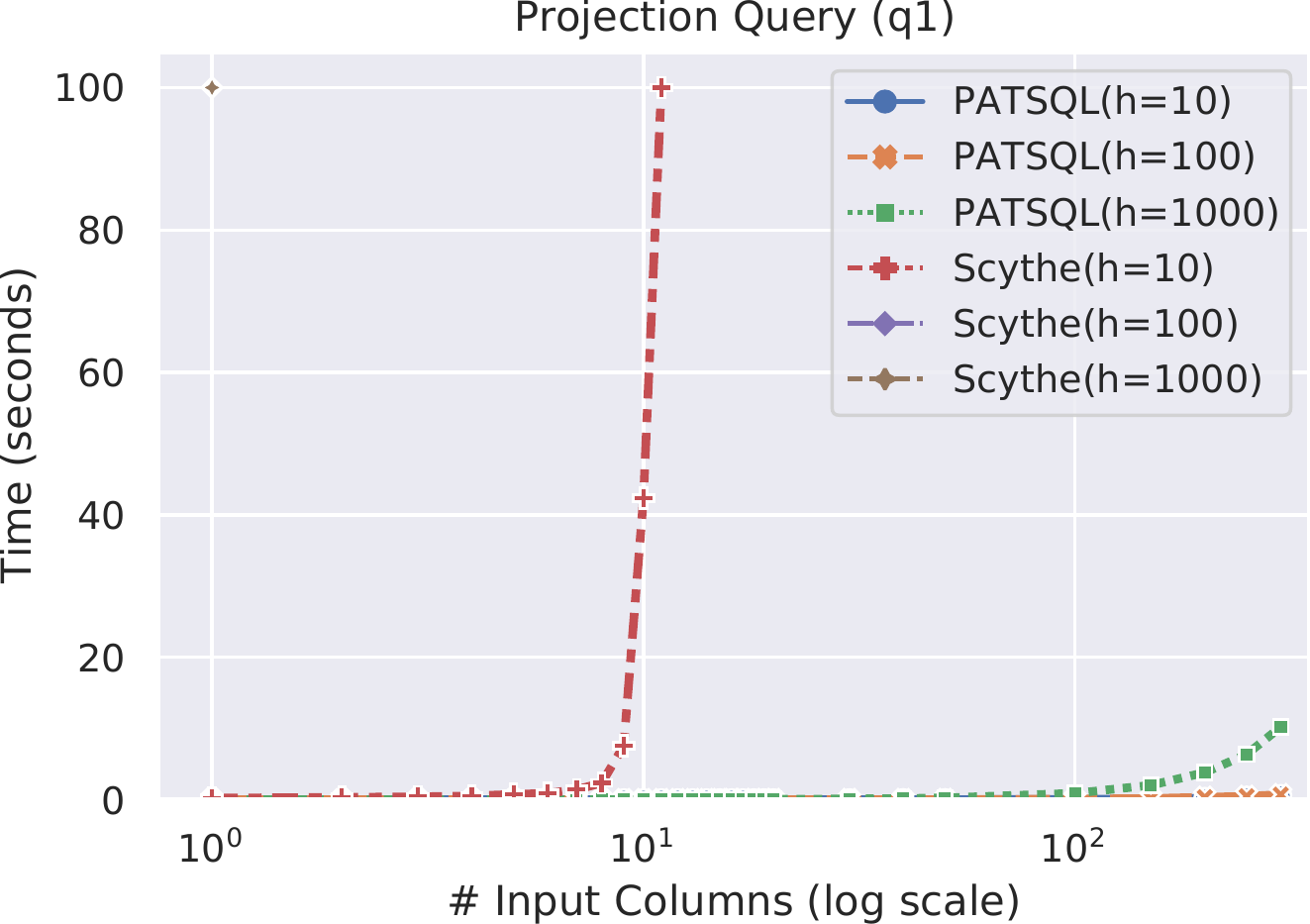}
    \includegraphics[width=.33\linewidth]{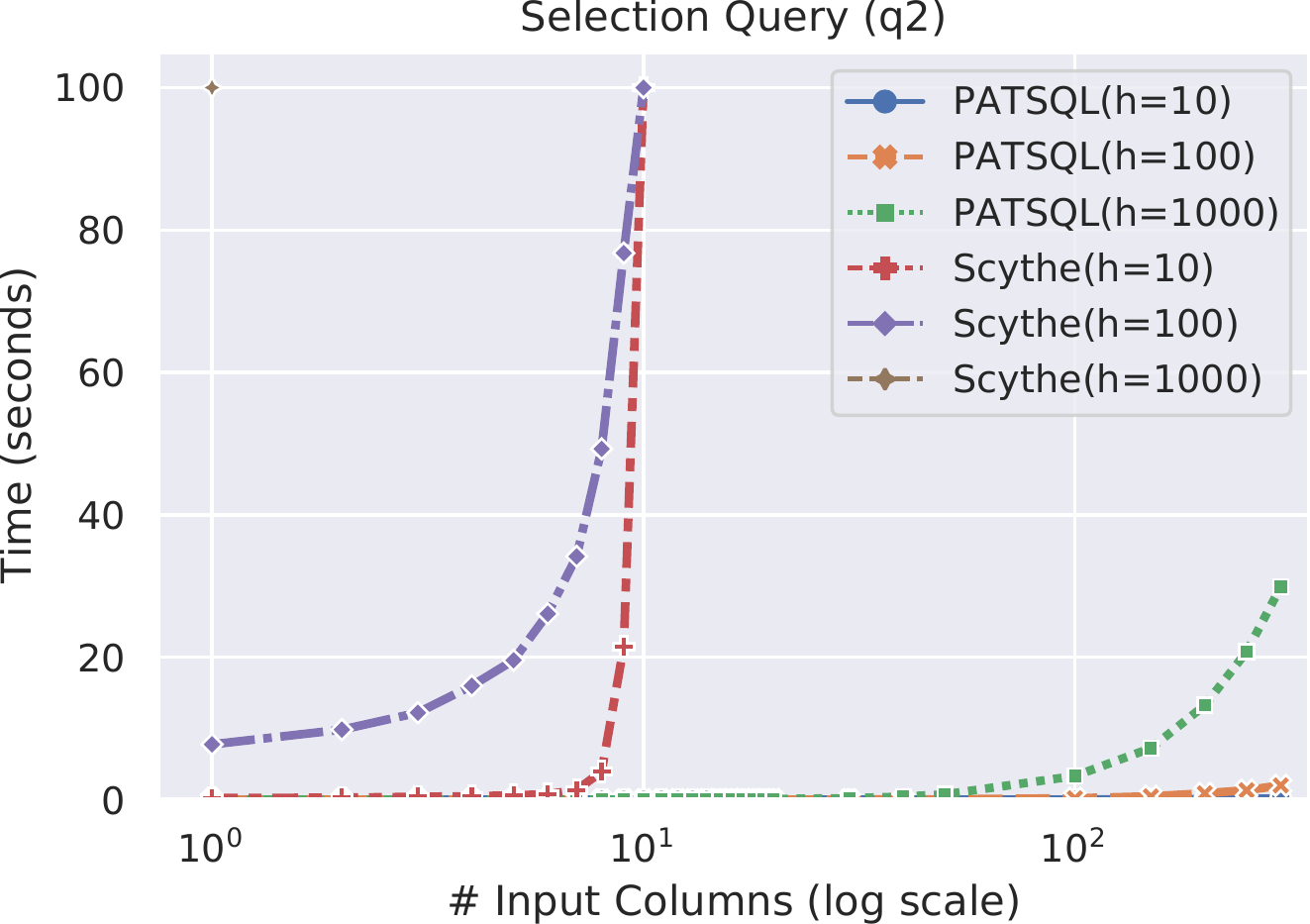}
    \includegraphics[width=.33\linewidth]{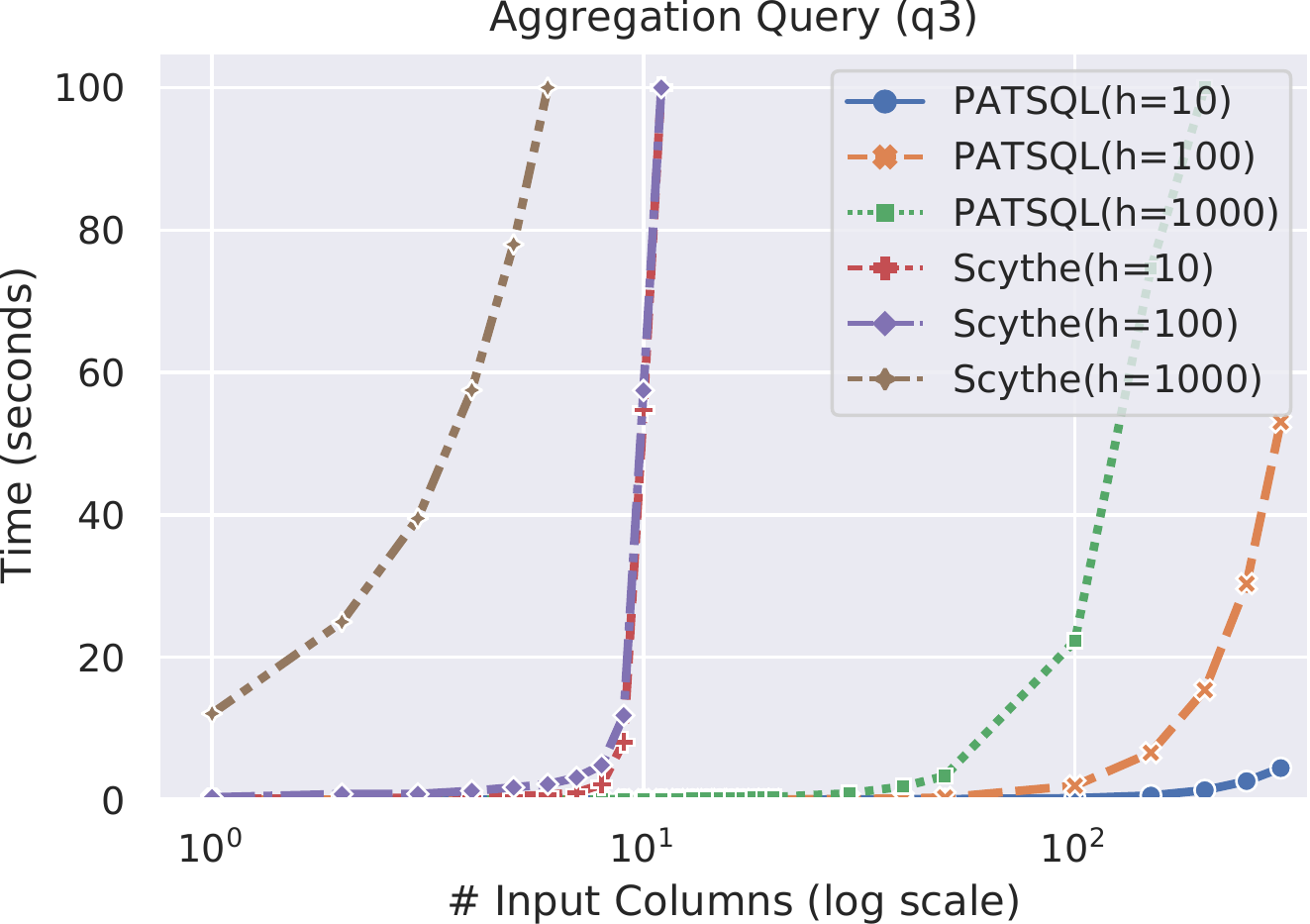}
    \caption{\added{The synthesis time in a controlled scalability experiment. "h"s in legends represent the number of input rows. For instance, "(h=10)" means that the number of input rows is 10.}}
    \label{fig:controlled_exp}
\end{figure*}%

\subsection{Comparison to Baseline}
To illustrate the effectiveness of our improvements in the algorithm, we compare the number of benchmarks solved by \tool\ and \textsc{BaseLine}. As a result, \textsc{BaseLine} solved \added{99} out of the 226 benchmarks while \tool\ solved \added{153} out of them (Table \ref{tab:result_benchmark}). 
Recall that the difference between \tool\ and \textsc{BaseLine}\ is in the completion of \texttt{Project} sketches. Therefore, the failed benchmarks by \textsc{BaseLine} were caused by inefficient performance in the completion of \texttt{Project} sketches. That is, the completion algorithm seen in prior work significantly decreases the performance in cases with large tables, and our improvements can \minor{work efficiently in} such cases. This difference leads to a \added{24\% $\left(= \left(153-99 \right)/226 \right)$} improvement in the number of solved benchmarks. 
Next, we compare the execution time. As a result, \textsc{BaseLine} solved \added{67} benchmarks within a second while \tool\ solved \added{136} within a second (Figure \ref{fig:time_benchmark}). The result shows that our improvements significantly reduce the execution time.
\minor{
Note that other improvements such as pruning by the table constraint $\varphi$ and the restriction on sketch structures in Table \ref{tab:comb} are not crucial for the performance improvement, but they are required to realize the efficient completion of \texttt{Project} sketches. 
}

For \textbf{RQ2}, the experimental result shows that our improvements in the algorithm improve \minor{the overall performance} significantly. We also show that the efficient completion of \texttt{Project} sketches is crucial for the overall performance. 

\subsection{Failed Cases}
To answer \textbf{RQ3}, we analyzed the \added{73} cases that \tool\ failed to solve. We classified the failure causes into two categories: \textit{f-elem} and \textit{f-struct}, and we found that \added{60} cases are in f-elem and \added{13} cases in f-struct (Table \ref{tab:failed_cases}). The `f-elem' row represents the number of queries that \tool\ failed to synthesize \minor{as the syntax features are not supported as of now.} \added{These include pivot operations (7 cases), \texttt{UNION} (6 cases), timestamp operations (5 cases), non equi-join (4 cases), \texttt{LIKE} (3 cases), \texttt{CASE} (3 case) and other 15 SQL features.} In contrast, the 'f-struct' row represents the number of failed cases due to the complex structures of desired queries. In particular, two cases in \texttt{kaggle} benchmark require sketches whose sizes are five, while the other cases require more than five. \tool\ cannot find relatively small sketches for \texttt{kaggle} benchmark because it needs to process larger I/O tables during the synthesis. Although the computational cost of \tool\ does not increase significantly with respect to the scale of I/O tables, it certainly does increase in a polynomial order. Therefore, \tool\ could not search a sufficient number of sketches before a timeout occurred. 

\added{
\subsection{Controlled Scalability Experiment}
To answer \textbf{RQ4}, we performed a controlled experiment that examines the difference between \tool\ and \scythe\ in synthesis time when the size of I/O tables increases. We used the following SQL queries as synthesis solutions in this experiment. 
}
\begin{enumerate}
    \item[(q1)] \verb|SELECT * FROM table;|
    \item[(q2)] \verb|SELECT * FROM table WHERE c1 = 'T';| 
    \item[(q3)] \verb|SELECT COUNT(*) FROM table;| 
\end{enumerate}
\added{
The queries q1, q2 and q3 represent the simplest queries for projection, selection and aggregation, respectively. We then crafted different I/O tables consisting of unique values to synthesize each of the queries. We employed different settings for the size of I/O tables. Specifically, the number of rows can be 10, 100 or 1,000 while the number of columns can be 1 to 300. For each pair of row and column sizes, we measured synthesis time taken to find the solution by \tool\ and \scythe. The timeout is set to be 100 seconds as in the other experiments. 
}

\added{Figure~\ref{fig:controlled_exp} shows the result. For the projection query q1, the time taken by \tool\ grows gradually as the column size increases. This is the case even when the number of rows is larger, such as 1,000 rows. In contrast, \scythe\ has difficulty finding a solution when the column size is 10 or more even though the number of rows is 10. \scythe\ was unable to find any solution for the case of 100 and 1,000 rows. For the other two queries q2 and q3, \tool\ was able to handle a large number of rows and columns that \scythe\ was not able to handle. We also found that the time taken by \tool\ gradually increased with the column size while that of \scythe\ increased significantly. In summary, this experiment shows \tool's strength in the scalability of I/O tables.}

\section{Related Work}
\textbf{Programming by Example}\ 
(PBE) is a technique that synthesizes programs from given I/O examples, and has been studied intensively in recent years \cite{microsoft-survey, pbe, ml4pbe}. PBE has been applied to help non-experts in a wide range of domains such as 
string manipulation \cite{flashfill, robustfill}, 
data migration \cite{mitra, dynamite}, 
data extraction \cite{flashextract} and 
MapReduce program \cite{map_reduce}. 
Several studies have proposed techniques that synthesize expressive SQL queries from I/O tables \cite{sql-synthesizer, scythe, squares, sqlsol}. 
SQLSynthesizer~\cite{sql-synthesizer} employs a kind of the decision tree algorithm to construct appropriate predicates in the \texttt{WHERE} clause. 
SqlSol~\cite{sqlsol} uses an off-the-shelf SMT solver to build the entire query by encoding SQL components and tables into logic constraints. 
SQLSynthesizer and SqlSol differ from our technique in that they do not support nested subqueries. 
\scythe~\cite{scythe} enumerates abstract queries in a bottom-up manner and instantiates each of them by encoding tables in bit-vectors. 
SQUARES~\cite{squares} is an SQL synthesizer developed on top of a state-of-the-art synthesis framework Trinity~\cite{trinty}. \scythe\ and SQUARES have similar performance on small examples. SQUARES performs better on tables having a large number of rows than \scythe, but it requires more types of hints including aggregation functions and attribute names. 
\minor{We cannot expect from a non-expert user to provide millions of rows in order to know how to write a correct query.} Our tool \tool\ significantly outperforms \scythe, as evidenced in Section \ref{sec:eval}, by utilizing properties in relational algebra. 
Our algorithm is also closely related to techniques that deal with table structures as I/O examples such as 
data frame manipulation \cite{morpheus, autopandas}, 
tensor manipulation \cite{tf_coder} and
data visualization \cite{vis_by}. 
These algorithms focus on methods that leverage the properties of table structures to enable efficient synthesis. Examples of them include pruning by table inclusion relations \cite{scythe, vis_by}, pruning by constraints on table metadata such as the number of columns and rows \cite{morpheus} and machine learning by transforming a table into a graph structure \cite{autopandas}. Also, many of them adopt the concept of sketch, which determines and validates the program structure before enumerating concrete programs, as our algorithm does. Thanks to these improvements, a rich set of syntax features and complex program structures have been supported. 
However, these existing techniques suffer exponential increases in computational cost as the number of columns increases. This challenge makes it difficult to apply the PBE methods in practical scenarios. Note that, to avoid such combinatorial explosions, some algorithms \cite{autopandas, sql-synthesizer, squares} are based on the assumption that the corresponding column names are always the same, which we believe \minor{is not practical especially when the user is not an expert.} \minor{To address these issues, \tool\ deals with I/O tables having full-scale schemas with no constraints on the column names.}
\vspace{1mm}
\\
\textbf{Query by Example} 
(QBE) is a technique for formulating database queries from several examples of records (and counterexamples in some cases) that the user wants to retrieve from tables \cite{squid}. QBE has been actively studied, and a variety of techniques have been proposed \cite{requery, fast_qre, qre_agg, squid}. In a typical setting for QBE, the target table is an existing one in a database, and therefore executing a query may take long time when the table has a large number of records. \minor{Thus, these algorithms need to deal with the scalability as the number of rows in the table increases, which our algorithm does not need to focus on.} Also, some techniques assume that helpful information such as used constants or used tables is not available. Despite these strict limitations, SQ{\footnotesize U}ID \cite{squid} restores queries with aggregation and grouping by pre-computeing statistics of semantic properties to construct an abduction-ready database. Although a rich set of syntax features has been supported, there are no QBE methods that support more advanced features such as nested query or window functions as far as we know. 
\vspace{1mm}
\\
\textbf{Synthesis Algorithms.}\ 
A wide variety of improvements in program synthesis algorithm have been proposed. Our algorithm is categorized as an enumerative synthesis algorithm, which is one of the most successful strategies \cite{microsoft-survey}. Smith et al. \cite{equivalence} proposed a program synthesis technique that only enumerates programs in \textit{normal form} to reduce the search space. The normal forms are computed from a set of rewrite rules on program structures. Although this concept is similar to our restriction on sketch structures in Table \ref{tab:comb}, we focus on restricting program structures for enabling efficient sketch completion, rather than reducing the search space on program structures.
We also highlight recent advances in synthesis techniques that employ machine learning or stochastic models \cite{ml4pbe, bigcode, deepcoder, infer_sketch, neural-synth, accelarate}. For example, S{\footnotesize KETCH}A{\footnotesize DAPT} \cite{infer_sketch} prioritizes sketches that may lead to a desired query based on a recurrent neural network model, which has learned patterns from codebases. \minor{Introducing such sketch priorities} into our algorithm would be straightforward since our sketch completion depends only on the target sketch.
\vspace{1mm}
\\
\textbf{Natural Language Interface}\ 
(NLI) is another approach than PBE for helping non-experts to write queries \cite{nli_db}. In particular, a variety of techniques for translating specifications in natural language into SQL queries have been studied \cite{SyntaxSQLNet, PointSQL, nli_schema_graph, IRNet}. For example, SyntaxSQLNet \cite{SyntaxSQLNet} is a text-to-SQL generator that employs an SQL specific syntax tree-based decoder. It supports a rich set of SQL features such as aggregation, union and nested query. However, specifications in natural language tend to be ambiguous compared to I/O examples. To address this issue, \textsc{Duoquest} \cite{duoquest} aims to combine PBE and NLI approaches to meet user's practical demands. We believe each advance in PBE and NLI technologies will also contribute to advances in such dual-specification approaches.

\section{Conclusion}
We have presented an SQL synthesizer called \tool, which synthesizes expressive SQL queries from input and output tables. \tool\ is the first SQL synthesizer that integrates properties known in relational algebra into sketch-based program synthesis. \tool\ employs a novel form of constraints and its top-down propagation mechanism for efficient sketch completion. We have shown that \tool\ outperforms a state-of-the-art algorithm \scythe\ in both the execution time and the scalability of input and output tables even though \tool\ does not employ hints about aggregation functions required by \scythe. 

An immediate future direction is to develop synthesis algorithms that support additional syntax elements and are able to find more complex program structures such as advanced SQL queries seen in Kaggle's tutorial. 
\added{ 
Another direction is to work with I/O examples that may contain incorrect values. In recent years, the program synthesis from such noisy data has been studied in several domains~\cite{robustfill, noisy}. 
}
It will also be interesting to consider the performance of synthesized queries although the performance can depend on the database indexes tuned for target tables.
More broadly, we believe that it is important to find out the requirements for the practical use of PBE tools through user studies and continue to improve the algorithms and user interfaces.

\begin{acks}
This work was supported by JSPS KAKENHI Grant Number\\
JP20H05706. 
\end{acks}

\balance

\bibliographystyle{ACM-Reference-Format}
\bibliography{ref}

\end{document}